\documentclass[journal=jpccck,manuscript=article,layout=twocolumn]{achemso}

\usepackage[utf8]{inputenc}
\usepackage[english]{babel}
\usepackage{graphicx}
\usepackage{dcolumn}
\usepackage{bm}
\usepackage{braket}
\usepackage{xcolor}
\usepackage{enumerate}
\usepackage{hyperref}
\usepackage{nicefrac}
\usepackage{hyperref}
\hypersetup{pdfborder=0 0 0,colorlinks=true,citecolor=blue,linkcolor=blue}
\usepackage[normalem]{ulem} 

\title{Tuning Einstein Oscillator Frequencies of Cation Rattlers: A Molecular Dynamics Study of the Lattice Thermal Conductivity of CsPbBr$_3$}

\author{Jonathan Lahnsteiner}
 \affiliation{VASP Software GmbH, Sensengasse 8/12, A-1090 Vienna, Austria}
\author{Max Rang}
\author{Menno Bokdam}\email{m.bokdam@utwente.nl}
 \affiliation{University of Twente, Faculty of Science and Technology and MESA+ Institute 
for Nanotechnology, P.O. Box 217, 7500 AE Enschede, The Netherlands}

\date{\today}
\begin{document}

\maketitle

\begin{tocentry}
\includegraphics[width=8.25cm]{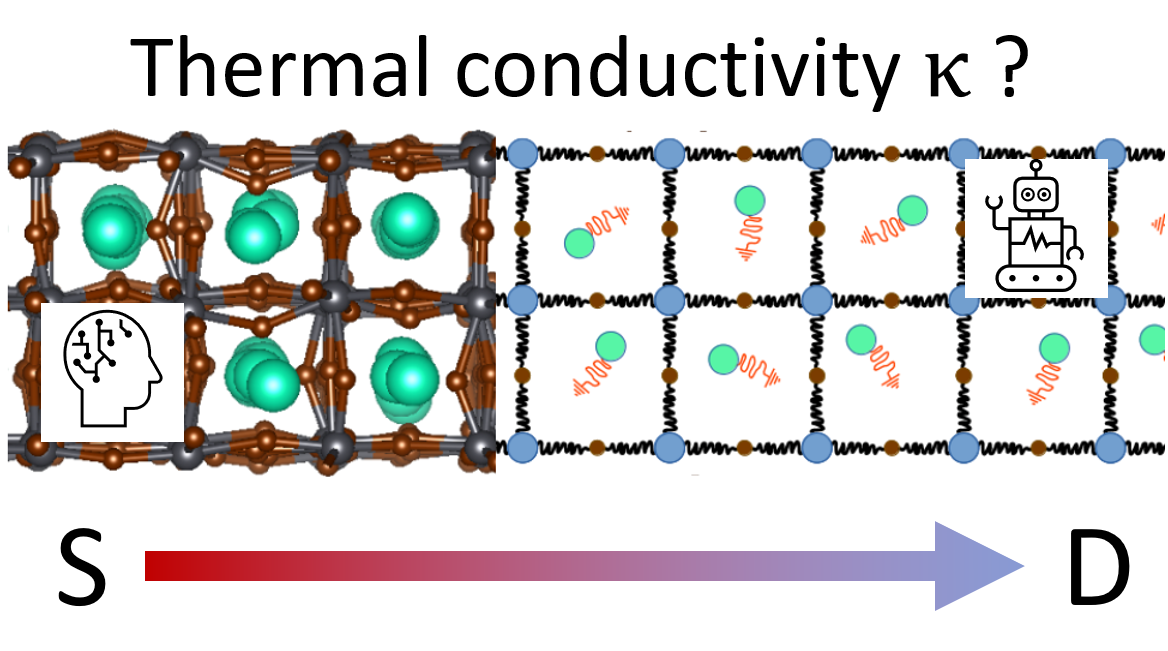}
\end{tocentry}

\begin{abstract}
The pure CsPbBr$_3$ perovskite is an  archetypal example of a strongly anharmonic crystal  that poses a major challenge for computational methods to describe its thermodynamic properties. Its lattice dynamics exhibits characteristics of a phonon liquid: mode coupling, low lifetimes and `rattlers'. To study the thermal conduction in this crystal, including the effect of dynamic disorder introduced by the Cs rattlers, we apply large-scale molecular dynamics (MD) simulations combined with machine-learning interatomic potentials. We simulate its ultra-low lattice thermal conductivity  in the cubic phase and obtain phonon spectra by measuring velocity autocorrelation functions. The thermal conductivity at 500~K is computed to be  $0.53\pm0.04\frac{\rm W}{\rm m K}$, which is similar to that of demineralized water under normal indoor conditions. MD based insight into the heat transport mechanism of halide perovskites is presented. In the analysis the Cs cations are interpreted as damped Einstein oscillators. The phonon bandstructure of a system with artificially raised Cs masses demonstrates an increased interference of the Cs rattling with the acoustic phonon modes. We show that  the thermal conductivity of the CsPbBr$_3$ perovskite can still be slightly decreased by tuning the cation rattling frequency into the range of the low lying acoustic modes.
\end{abstract}

\section{Introduction} 
\begin{figure}[b!]
 \centering
 \includegraphics[width=\columnwidth]{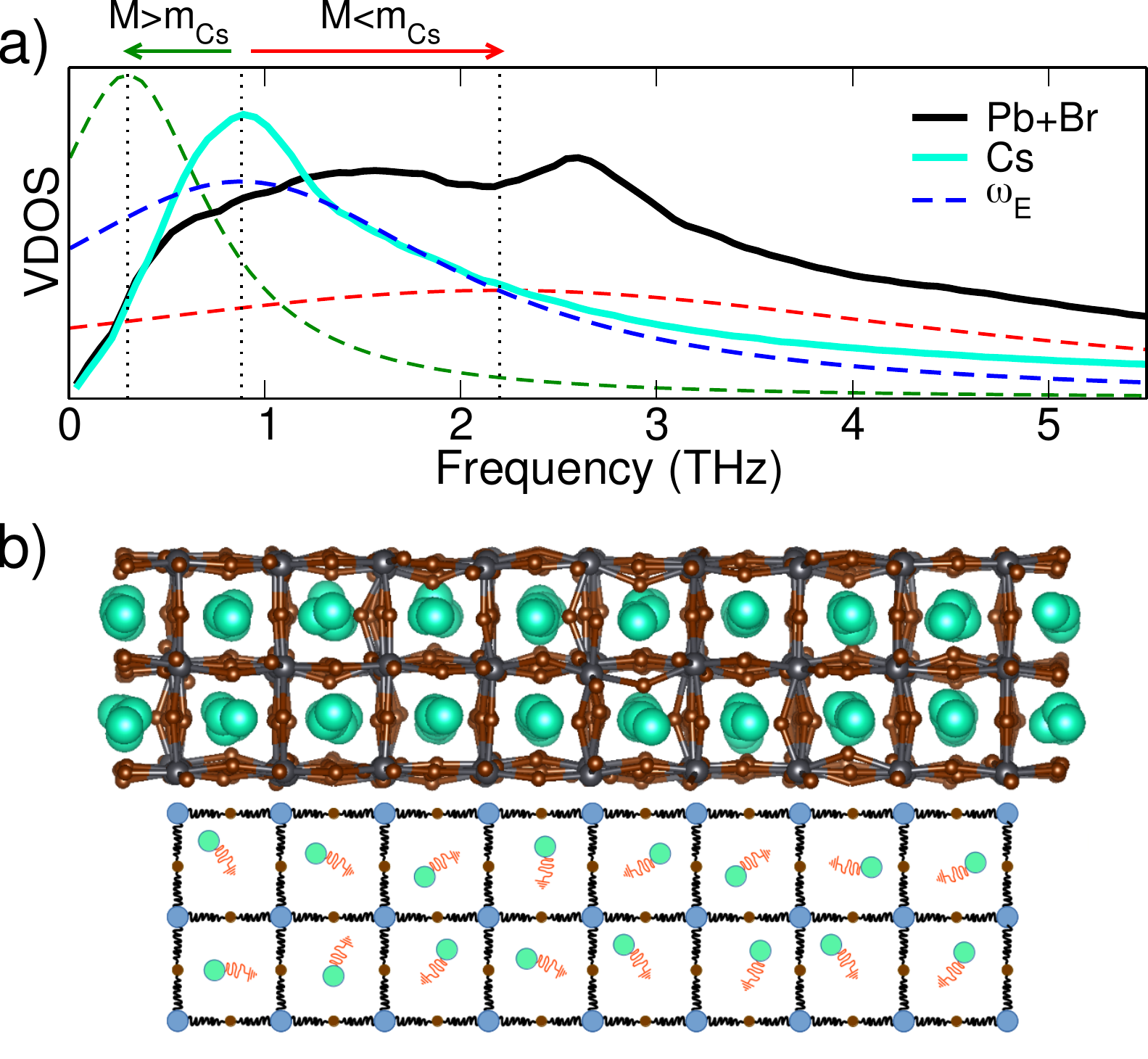}
 \caption{Cs rattlers as Einstein oscillators in CsPbBr$_3$. (a) The vibrational density of states (VDOS) of CsPbBr$_3$ at 500~K decomposed in a Pb+Br and a Cs contribution. A Lorenzian distribution (dashed lines) centered at frequency $\omega_{\rm E}$ approximates the Cs contribution of the VDOS. Changing the Cs mass can blue or red shift $\omega_{\rm E}$. (b) Opposite to Pb and Br, Cs atoms are approximated as not interconnected, damped oscillators in the PbBr$_3$ framework. }  \label{fig:1b}
\end{figure}

Around room temperature, condensed matter is often represented as a `wiggling and jiggling' of atoms connected by springs. When this matter displays periodicity, as in a crystal lattice, energy can be stored in collective lattice oscillations. The phonon gas model is a 0~K approximation that assumes that these springs are harmonic. The phonons in the lattice possess infinite lifetimes and can transport kinetic energy unhindered throughout the crystal. In reality, there is always a degree of \textit{anharmonicity} of the interatomic potential and \textit{disorder} in the atomic positions. This causes phonon modes to scatter frequently with one another, thereby losing their phase coherence, resulting in finite lifetimes. Even though present day simulations go beyond the harmonic approximation, much is still unknown about thermal transport in crystals that can be classified as strongly anharmonic\cite{Simoncelli:natp19}. The CsPbBr$_3$ perovskite is a clear example of a strongly anharmonic crystal\cite{Miyata:scadv17,Lanigan-Atkins:natm21}, and forms a challenge for first-principles simulation methods\cite{Simoncelli:natp19,Gehrmann:natc:19,Songvilay:prm19,Jinnouchi:prl19,Zhu:acsami22}. This type of perovskite is technologically interesting because of its opto-electronic effects~\cite{Eaton:pnas16,Chen:nanol17,Wang:nanol18} and its \textit{ultra-low} thermal conductivity ($\kappa<1\frac{\rm W}{\rm m K}$)~\cite{Yuping:com14,Mettan:jpcc15,Yue:prb16,Filippetti:jpcc16,Wang:nanol18}. The material is cheap and easy to produce, because it can be crystallized from its liquid solutions, but it is also toxic and not very stable~\cite{Wang:nanol18,Guo:acsel17}. Its perovskite lattice contains a Cs$^+$ `rattler'\cite{Miyata:scadv17,Lee:pnas17} in the center of the lead-bromide cage, which surrounds the cation as shown in Figure~\ref{fig:1b}. Neighbouring Cs cations show a close to negligible level of dynamic correlation. The effect of rattling has been linked to the low thermal conductivity of several (types of) halide perovskites\cite{Xing:sc13,Miyata:scadv17,Lee:pnas17,Xie:jacs20,Jong:chc22,Acharyya:natc22}, possibly inspired by resemblances with rattlers observed in host-guest complexes: filled skutterudites\cite{Hermann:prl03,Ge:aem22} or clathrates\cite{Chen:fier18}. Interest in these materials results from efforts to find a phonon-glass electron-crystal\cite{Takabatake:rmp14}, which would make an ideal thermoelectric. The effect of the guest rattlers is often described by localised, independent Einstein oscillators\cite{Hermann:prl03}. In this scenario these loosely bound, often heavy, atoms produce strong incoherent scattering that impedes thermal transport as illustrated in Ref.~\cite{Hermann:ajp04} and by molecular dynamics (MD) simulations in Ref.~\cite{Dong:prl01}. However, this explanation has been challenged by experimental work indicating that there is a strong coupling between guest and host modes \cite{Koza:natm08,Christensen:natm08,Toberer:jmc11}. Furthermore, simulations have indicated that it is the phonon lifetimes (over a broad spectral range) that are affected by the rattlers rather than their group velocities\cite{Wu:prb14,Tadano:prb15}. 
These differences arise because of uncertainties in the coupling strength between the host's phonons and the rattler phonons. These uncertainties are also present in the halide perovskites, which raises the question how rattling affects CsPbBr$_{3}$. In this work we simulate the lattice thermal transport of this perovskite in its cubic phase using an on-the-fly trained machine learning (ML) potential\cite{Jinnouchi:prl19,Lahnsteiner:prb22}. We study the influence of rattling by a computational experiment in which we tune the rattling frequency and the level of disorder of the Cs atoms.

\vspace{2mm}

The challenge in modeling thermal transport in strongly anharmonic crystals stems from the complexity of their potential energy surface (PES). 
As a result a highly interacting phonon gas, or maybe even phonon liquid, is formed in the cubic phase of CsPbBr$_3$\cite{Lahnsteiner:prb22}. Most of the phonon quasi-particles are ill-behaved; with non-Lorentzian multipeak spectra and/or lifetimes smaller than the inverse of their frequency (Ioffe-Regel limit). In CsPbBr$_{3}$ these two criteria are not satisfied, since the phonon spectra show mode coupling~\cite{Simoncelli:natp19,Lahnsteiner:prb22}. This implies that lattice thermal transport may not be accessible with models ($ \kappa_{\rm L}$) relying on the Peierls-Boltzmann (PB) transport equation in the relaxation time approximation,
\begin{equation}
 \kappa^{xy}_{\rm P}\propto \sum_{\mathbf{q},\alpha}C_\alpha(\mathbf{q})v^x_\alpha(\mathbf{q})v^y_\alpha(\mathbf{q})\tau_{\alpha}(\mathbf{q}). 
 \label{eq:P}
\end{equation}
PB relies on decoupled and well separated phonon branches  with frequencies $\omega_\alpha(\mathbf{q})$ and group velocity $\mathbf{v}_\alpha{}(\mathbf{q})=\nabla_\mathbf{q}\omega_\alpha(\mathbf{q})$. Each phonon mode $\alpha$ has a relaxation time $\tau_\alpha$, a population $n$ and contributes $C_\alpha(\mathbf{q})=\frac{(\hbar\omega_\alpha(\mathbf{q}))^2}{k_{\rm B}T^2}n(\omega_\alpha(\mathbf{q}))\left(n(\omega_\alpha(\mathbf{q}))+1\right)$ to the lattice heat capacity.

The phonon dispersion $\omega_\alpha(\mathbf{q})$ can be readily obtained from standard \textit{ab-initio} calculations, by computing the second order (harmonic) force constants. However, higher order potential derivatives are required to obtain finite lifetimes\cite{Hellman:prb13A,Reissland:Book1973,Allen:prb10,Cowley:ap63}. These calculations are exceedingly expensive and can quickly become computationally intractable for complex materials, but have been done for CsPbBr$_3$ including effects of up to 3rd and/or 4th order derivatives\cite{Simoncelli:natp19,Tadano:prl22,Wang:prb23}.

The work of Simoncelli~\textit{et al.}\cite{Simoncelli:natp19} goes beyond PB transport and should be better at describing thermal transport in anharmonic lattices. To summarize, they approximate the lattice thermal conductivity as the sum of two contributions, 
\begin{equation}
 \kappa_{\rm L}=\kappa_{\rm P}+\kappa_{\rm C}.
\end{equation}
This includes a Peierls (populations) term corresponding to \textit{intra}band transport ($\kappa_{\rm P}$ in Eq.~(\ref{eq:P})) and a coherences term corresponding to \textit{inter}band transport ($\kappa_{\rm C}$). These contributions scale differently with the lifetime; with decreasing $\tau$, $\kappa_{\rm P}$ reduces, whereas $\kappa_{\rm C}$ increases. The latter describes diffusive heat transport carried by diffusons as identified by Allen and Feldman\cite{Allen:pmb99}. Applying the thermal conductivity equation of Simoncelli \textit{et al.} to the case of low temperature (50-300~K), orthorhombic CsPbBr$_3$ shows a temperature dependent thermal conductivity $\kappa_{\rm L}(T)$ in agreement with experiment\cite{Simoncelli:natp19}. Recently, Tadano~\textit{et al.}\cite{Tadano:prl22} have applied the same formalism, but used renormalized frequencies $\{\Omega_\alpha(\mathbf{q})\}$ originating from a $GW$-type approach that describes the broadening of the quasi-particle spectrum by phonon-phonon interactions. They find, in the high temperature (400-800~K) cubic phase, a small temperature dependence  with a maximum  $\kappa_{\rm L}(\sim450{\rm K})=0.51\frac{\rm W}{\rm m K}$, and $\kappa_{\rm L}(500{\rm K})=0.50\frac{\rm W}{\rm m K}$. Supporting findings have been very recently reported by Wang~\textit{et al.}\cite{Wang:prb23}. However, depending on the level of self-consistent phonon theory to calculate $\{\Omega_\alpha(\mathbf{q}),\tau_\alpha(\mathbf{q})\}$, differences in $\kappa_{\rm L}$ of up to 50\% are reported.

\vspace{2mm}

We make use of recent developments in ML interatomic potentials\cite{Behler:prl07,Bartok:prl10,Rupp:prl12,Bartok:prb13} to simulate heat transport using the full PES and (in principle) including all higher order scattering processes. These developments have opened up a computationally efficient description of the interatomic potential. Using MD and this potential we simulate large thermodynamic ensembles of cubic CsPbBr$_3$ at 500~K with near-first principles accuracy. Properties related to ion dynamics, such as thermal conductivity or phonon dispersion are extracted by a `measurement' of the temperature gradient or the projected velocity auto-correlation functions. This is a purely classical approach, valid above the Debye temperature, that naturally includes the effect of dynamical disorder that is absent or approximated in the before mentioned lattice thermal conductivity models. Disorder is expected to drive the system further away from the harmonic regime. The heat transport coefficient ($\kappa$) is calculated by  thermal conductivity measurements in a non-equilibrium MD (NEMD) setup\cite{Mueller:jcp97,Hafskjod:mp93,Dunn:jap16,Li:jcp19,Zhou:prb19}. As with the lattice thermal conductivity models, this method has its own convergence conditions. Using the trained ML potential, we here observe a relatively fast convergence of $\kappa$ with MD simulation size/length. 
However, we note that the required simulation size is too large for a fully \textit{ab-initio} MD to be computationally feasible. In order to increase our phenomenological understanding of heat transport in this system, we compare to the predictions made by the lattice models\cite{Simoncelli:natp19,Tadano:prl22,Wang:prb23}. we probe phonon frequencies, lifetimes and populations under full phonon-phonon scattering conditions in the microcanonical ensemble.  Even though the interband contribution to the thermal transport may be substantial, we show that the remaining Peierls (intraband) contribution is affected by Cs rattling. Therefore there is room for changing the thermal conductivity by tuning the effective rattling frequency.

\vspace{2mm}

\begin{figure}[t!]
 \centering
 \includegraphics[width=\columnwidth]{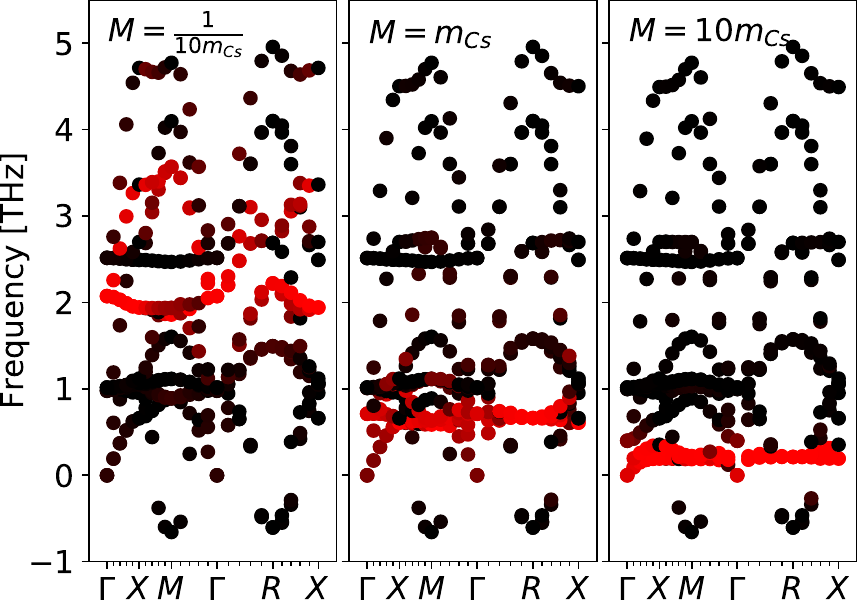}
 \caption{Shifting the 'rattling band'. Harmonic phonon bandstructure of cubic CsPbBr$_3$ with \textit{light} ($\frac{m_{\rm Cs}}{10}$), \textit{normal} ($m_{\rm Cs}$) and \textit{heavy} ($10m_{\rm Cs}$) effective mass of the Cs cation. The colorscale from black to red indicates the Cs contribution to the respective eigenmode.} 
\label{fig:1}
\end{figure}

\section{Computational experiment}
In Figure~\ref{fig:1} the harmonic phonon dispersion\cite{Landau:StatPhys1,Togo:phonopy15} $\omega_\alpha(\mathbf{q})$  of cubic CsPbBr$_3$ is shown. By highlighting (in red color) the Cs contribution in the respective eigenmode $\frac{|e_{{\rm Cs},q}(\omega)|^2}{|\mathbf{e}_{q}(\omega)|^2}$, we identify nearly dispersionless `\textit{rattling bands}'. 
In the middle panel we see that the Cs cations mainly contribute to the phonon bandstructure in an energy window around $\sim$0.7~THz. In the left/right panel we have artificially lowered/raised the masses of only the Cs atoms, which results in a blue/red shift of the central rattling frequency, respectively. For the light Cs cations the central frequency shifts up to $\sim 2.2$~THz, thereby almost removing the overlap with the linear acoustic bands. Conversely, the rattling band of the heavy Cs cations now overlaps at $\sim 0.2$~THz with the acoustic bands. Here, we will explore the effects of this rattling band on the lattice thermal conductivity of CsPbBr$_3$. In the analysis we view the Cs rattlers as Einstein oscillators with a characteristic frequency ($\omega_{\rm E}$) and a Lorentzian distribution in the vibrational density of states (VDOS) as plotted in Figure~\ref{fig:1b}. We perform MD simulations for different Cs masses ($M$) to study the influence of rattlers on the phonon spectrum and the thermal conductivity. Altering the mass does not change the PES as described by Density Functional Theory (DFT), but modifies the Einstein frequency as $\omega_{\rm E}=\sqrt{\frac{k_{\rm E}}{M}}$, with $k_{\rm E}$ the spring constant. Furthermore, it affects the level of disorder exhibited by the Cs atoms.  The same perturbation in the force has a larger effect on the displacement of a light compared to a heavy Cs atom. Only by performing the MD simulation can we find out how these two effects affect $\kappa$. Therefore, for the systems with a low, normal and high Cs mass, we apply two types of simulations: \textit{i}) a passive measurement of the phonon spectrum using VACFs, and \textit{ii}) an active measurement of the thermal conductivity.

\section{Computational Method}

We start by briefly summarizing the applied computational approach as a step-by-step plan:
\begin{itemize}
    \item Train MLFFs for CsPbBr$_3$ with light, normal and heavy Cs masses.
\end{itemize}
Then, for each of the three different mass systems we: 
\begin{itemize}
    \item Simulate $NVE$ ensembles in a $10\times10\times10$ cubic supercells of CsPbBr$_3$ at 500~K.
    \item Calculate $\mathbf{q}$-VACFs in post-processing of the $NVE$-trajectories for all phonon modes. Calculate PVACFs for the first acoustic mode and fit Lorentzians to the power spectrum.
    \item Simulate $NPT$ ensembles of $4\times4\times20$ supercells at 500~K and 1 bar. 
    \item Simulate heat flow in a NEMD setup of $4\times4\times20$ supercells with leads set at 500$\pm25$~K. Five independent trajectories starting from snapshots of the $NPT$ ensemble are calculated.
    \item The ensemble average temperature gradient and heat flux are computed in post-processing and result in the thermal conductivity.
\end{itemize}

We now expand on some of the most important computational details of these steps. More details have added to the supplementary information (SI).

We simulate the heat flow through the lattice using large-scale MD and the Machine-Learning Force-Fields (MLFF) method \cite{Jinnouchi:prl19,Jinnouchi:prb19,Jinnouchi:jcp20} as implemented in the DFT code \textsc{vasp}\cite{vasp-1,vasp-2} version~6.3. The MLFF training procedure and its accuracy benchmarking are described in the SI. Here, we want to mention that the MLFFs were trained with respect to the PES as described by DFT using the SCAN\cite{Sun:prl15} exchange-correlation functional, which has been shown to be accurate for this type of perovskite\cite{Bokdam:prl17,Lahnsteiner:prm18}.

The $\mathbf{q}$-resolved VACFs~\cite{Sun:prb10,Zhang:prb17,Sun:prb14,Lahnsteiner:prb22} are computed from microcanonical ensembles. The velocity field of atoms type $s$ in reciprocal space is given by
\begin{equation}
    \mathbf{V}_s(\mathbf{q},t) = \sum_{\mathbf{n}} \sqrt{m_s} \mathbf{v}_s (\mathbf{n},t) e^{i\mathbf{q}\cdot \mathbf{r}_s(\mathbf{n},t)}.
\end{equation}
Here $\mathbf{v}_s(\mathbf{n},t)$ is the velocity at position $\mathbf{r}_s(\mathbf{n},t)$ at time $t$ inside 
the unit cell denoted by $\mathbf{n}$ with mass $m_s$. We compute the self-correlation and perform a
temporal Fourier transform, i.e.
\begin{equation}
    H_s(\mathbf{q},\omega) = \int \mathbf{V}_s(\mathbf{q},t')\mathbf{V}_s(-\mathbf{q},t'')e^{-i\omega t}dt,
    \label{eq:3}
\end{equation}
where $\omega$ denotes the temporal frequency axis and $t=t'-t''$. By taking the power spectrum of this quantity the $\mathbf{q}$-resolved form of the phonon density of states (DOS) is obtained. The projected VACF is obtained by projecting $\mathbf{v}_s (\mathbf{n},t)$ along the harmonic phonon eigenvectors $\mathbf{e}_{s,\alpha}(\mathbf{q})$, summing over the species index $s$, and computing the transform as in Eq.~(\ref{eq:3}). The resulting PVACF $H_\alpha(\mathbf{q},\omega)$ is then decomposed into 15 phonon branches  $\alpha$ of the CsPbBr$_3$ cubic phase. The (P)VACFs were computed with the open-source code \textsc{dsleap}\cite{Lahnsteiner:prb22,DSLEAP22}. 

To measure the thermal conductivity in NEMD,  we apply the approach of a constant temperature gradient ($\frac{\Delta{}T}{\rm L}$)~\cite{Dunn:jap16,Li:jcp19,Zhou:prb19}, imposed by the Langevin thermostat~\cite{Bussi:pre07} with a low friction coefficient of 1~ps$^{-1}$. We observe that this approach converges the thermal conductivity value within acceptable simulation times, see SI. The geometry is split into three regions: a hot region ($\Gamma_{\rm H}$), a cold region ($\Gamma_{\rm C}$) 
and a scattering region of length $\rm L$  where no thermostating is applied. The desired temperatures for $\Gamma_{\rm H}$ and $\Gamma_{\rm C}$ are pinned to $\rm T_{H}$ and $\rm T_{C}$ by the thermostat. As a result, an energy flux

\begin{equation}
    \Delta Q(t) = \frac{1}{\rm A \Delta t}\left[E_{\rm hot}(t)-E_{\rm cold}(t)\right],
\end{equation}
in opposite direction of the temperature gradient is induced. Here, $E_{\rm hot}$ and $E_{\rm cold}$ are the amounts of kinetic energy the thermostats insert into and extract from the simulation cell in one time step $\Delta t$, respectively, and $\rm A$ is the area perpendicular to the heat flow. To compute the thermal conductivity Fourier's law is used; 
\begin{equation}
	\langle\Delta Q\rangle = -2\kappa \langle\nabla{}T\rangle,
	\label{heatkappa}
\end{equation}
where $\langle.\rangle$ denote ensemble averages. The factor two arises due to the applied periodic boundary conditions, resulting in two scattering regions in each simulation cell (see Fig.~\ref{fig:TGrad}). Therefore, mirror symmetry is applied in determining the average temperature profile.

\begin{figure}[t]
   \centering
   \includegraphics[width=1.0\columnwidth]{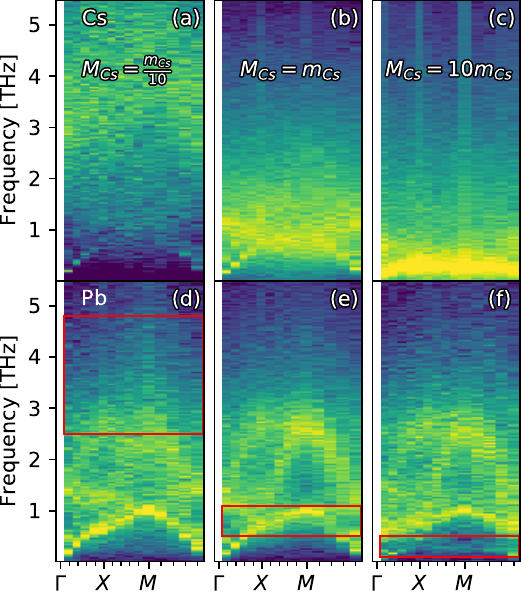}
   \caption{$\mathbf{q}$-VACF of Cs and Pb atoms for light, medium and heavy Cs masses. The frequency domain of the Cs rattling band is indicated by the red rectangles in (d-f).}
   \label{fig:CsPb}
\end{figure}

\begin{figure}[!t]
 \centering
 \includegraphics[width=\columnwidth]{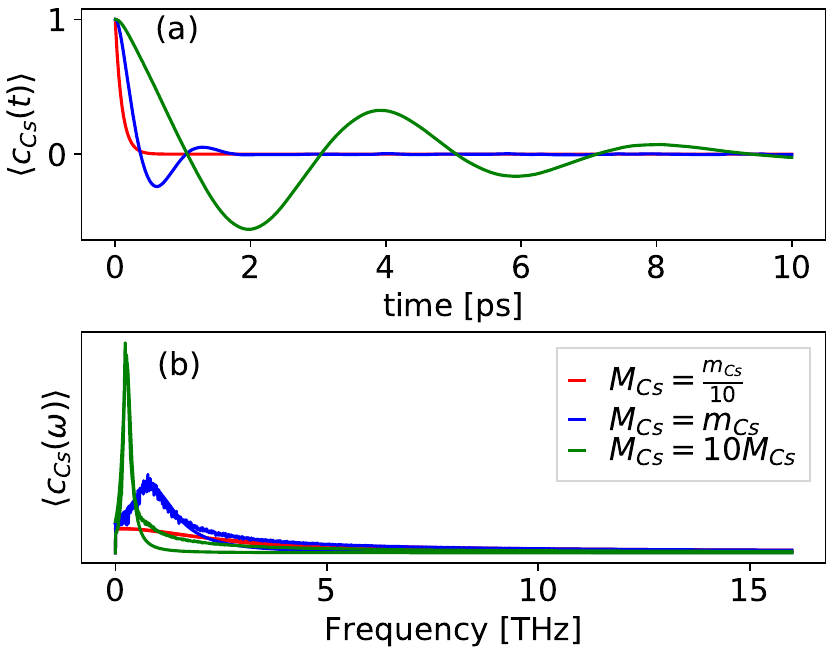}
	\caption{(a) Averaged Cs displacement self-correlation function and its (b) power spectrum  for light, normal and heavy Cs mass. The displacement vector connects the Cs atom and the center of the surrounding PbBr cage.} 
	  \label{fig:CsRattler}
\end{figure}
\section{Results}
We start with a characterization of the anharmonic vibrations in the three CsPbBr$_3$ model systems as they appear in MD. $\mathbf{q}$-resolved VACFs were computed from 10 microcanonical MD runs at $500$~K in $10\times10\times10$ cubic supercells. The averaged power spectrum of Eq.~(\ref{eq:3}) is shown in Figure~\ref{fig:CsPb} and shows the Cs and Pb contribution to the phonon spectrum. As in the harmonic approximation, decreasing the mass in the MD simulation shifts the Cs rattling band up, as can be seen in Fig.~\ref{fig:CsPb}(a). The shift is somewhat larger than in Fig.~\ref{fig:1}.
In Fig.~\ref{fig:CsPb}(c) we see that the rattling frequency is shifted down in the case of heavy Cs atoms. We are interested in their impact on Pb vibrations in the broad rattling window indicated by the red rectangles in Figs.~\ref{fig:CsPb}(d-f). We will get back to this point later on.

The rattling Cs atoms introduce dynamical disorder into the system, which we measure by the displacement $\mathbf{d}_{Cs}(t)$ of the Cs atom from the geometric center of its enclosing PbBr host lattice.
We extract effective rattling frequencies $\omega_{\rm E}$ and correlation times from the self-correlation functions $\langle C_{Cs}(t)\rangle =\langle \mathbf{d}_{Cs}(t)\mathbf{d}_{Cs}(0)\rangle$ as shown in Figure~\ref{fig:CsRattler}(a). The normal and heavy mass (blue and green) behave as \textit{damped} Einstein oscillators, the light mass shows \textit{overdamped} dynamics. All have very short self-correlation times, ranging from sub-picoseconds (low mass) to several picoseconds (heavy mass). The power spectra of the Cs motion shown in Fig.~\ref{fig:CsRattler}(b) show clear peaks in the case of normal and heavy Cs masses that are fitted with Lorentzians. These peaks are located at $\omega_{\rm E}=$ $0.8$~THz and $0.3$~THz, respectively. The lighter systems show broader Lorentzians, indicative of more dynamic disorder. The spectrum of the light mass system is so broad, that it more closely resembles white noise than a Lorentzian.  In this case the rattling picture does not apply anymore.\newline

\begin{figure}[!b]
   \centering
   \includegraphics[width=\columnwidth]{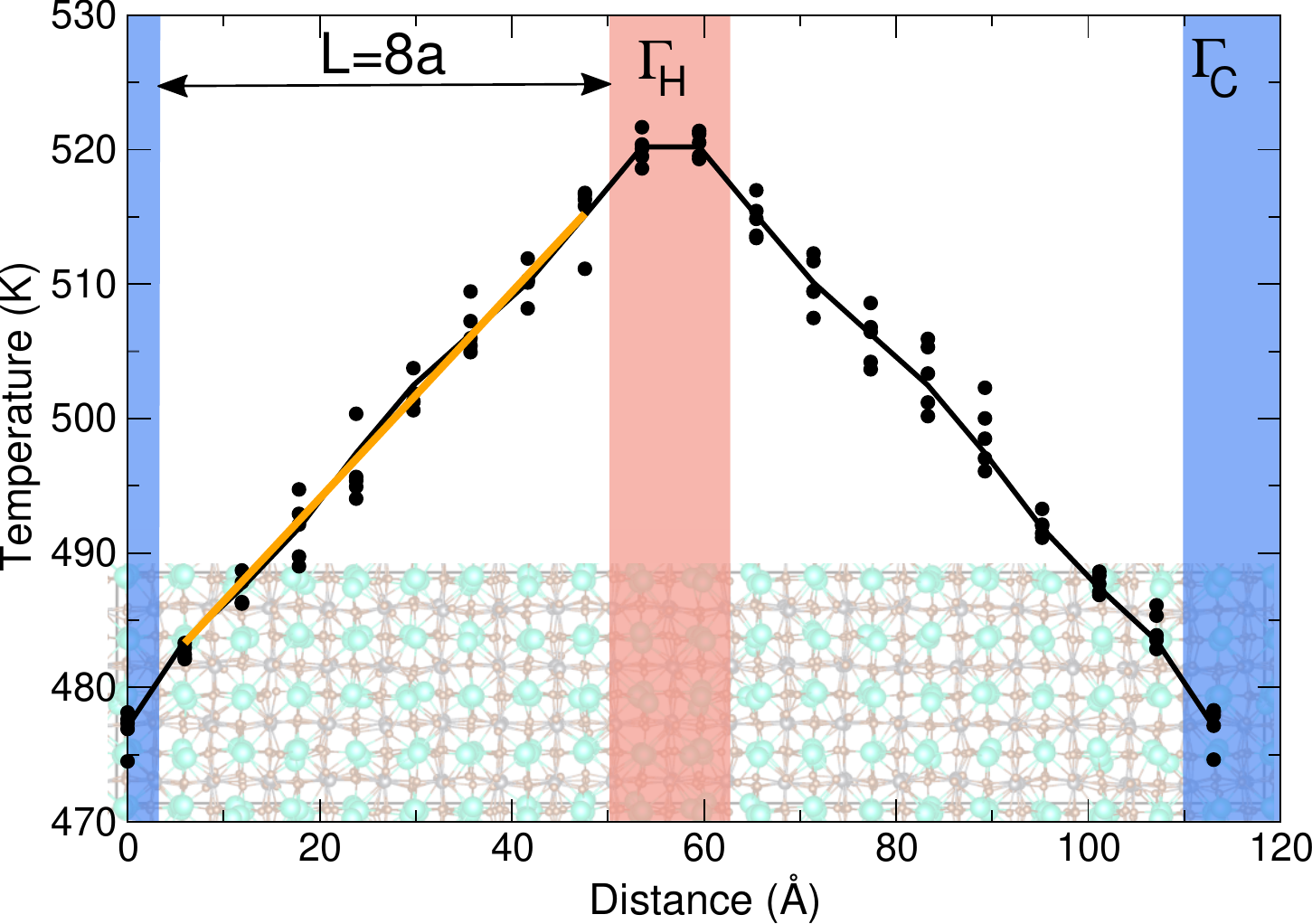}
\caption{Temperature profile obtained from five independent NEMD simulations of CsPbBr$_3$ (normal mass). Thermostats keep the (red) $\Gamma_{\rm H}$-layers at high and the (blue) $\Gamma_{\rm C}$-layers at low constant temperature. A stable temperature profile under periodic boundary conditions develops by time averaging the layer-resolved temperatures.  The orange line is the fit determining $\langle\nabla{}T\rangle$.}
   \label{fig:TGrad}
\end{figure}

\begin{figure*}
 \centering
 \includegraphics[width=2\columnwidth]{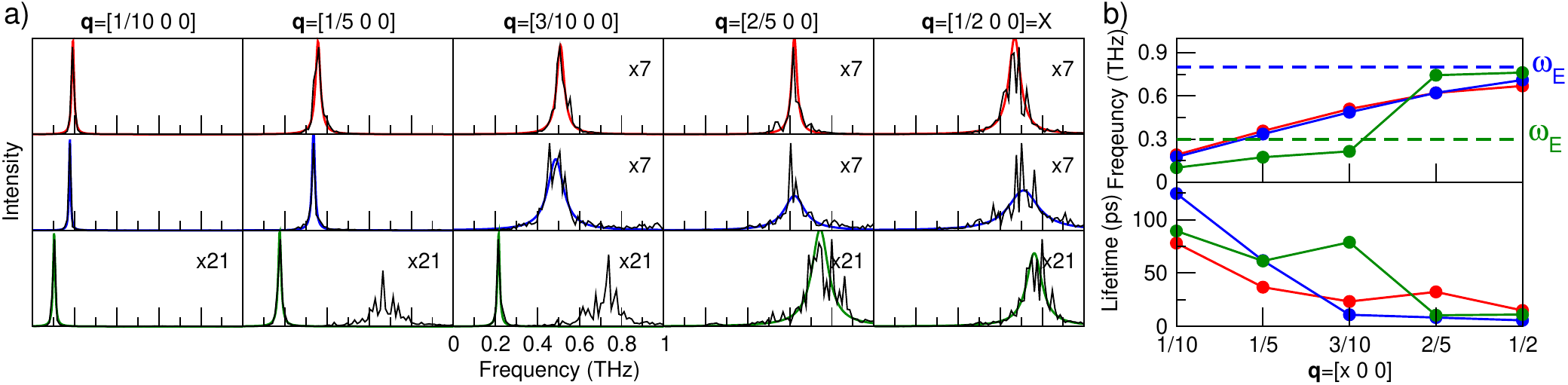}
\caption{Projection of the first acoustic mode ($\alpha=1$) on the MD trajectory. a) $\mathbf{q}$-PVACF spectra for $\mathbf{q}$-points on the $\Gamma-$X line. The spectra (black lines) for the low, normal and high Cs mass are shown from top to bottom. In some of the subplots the intensity of the spectra have been multiplied by the indicated factors. A Lorentzian function (color lines) is fitted to each spectrum. b) The frequencies and lifetimes of the fitted Lorentzians.} 
\label{fig:coupling}
\end{figure*}

\begin{table}[b!]
	\caption{Lattice thermal conductivity ($\kappa$) in $\frac{\rm W}{\rm m K}$ for CsPbBr$_3$ at 500~K with different Cs 	masses. Confidence intervals at 95\% uncertainty are computed with the Student's t test~\cite{student1908probable}.}
	\begin{tabular}{|m{0.35\columnwidth}|m{0.15\columnwidth}|m{0.15\columnwidth}|m{0.15\columnwidth}|m{0.15\columnwidth}|}\hline
		   &  light mass   & normal mass  &  heavy mass \\
	\hline
	Run 1 &  0.52 & 0.49 &  0.37  \\
	Run 2 &  0.53 & 0.48 &  0.36  \\
	Run 3 &  0.65 & 0.60 &  0.39  \\
	Run 4 &  0.59 & 0.57 &  0.36  \\
	Run 5 &  0.61 & 0.51 &  0.32  \\
	\hline
	Mean  &  0.58 & 0.53 &  0.36  \\
 	Confidence interval &$\pm$0.049 & $\pm$0.044 & $\pm$0.023 \\
    \hline
	\end{tabular}
	\label{tbl:1}
\end{table}

\noindent The lattice thermal conductivity of CsPbBr$_3$ in its cubic phase at 500~K is determined by computational NEMD experiment in which the average (over time) heat flux between two constant temperature leads and the established thermal gradient in the scattering region are measured. Figure~\ref{fig:TGrad} shows the layer-resolved temperature profile (black solid line) obtained from this simulation. The profile is the average over five trajectories of 1~ns each, starting from independent snapshots of a $NPT$-ensemble of the same system and temperature under atmospheric pressure. The temperature gradient is determined by a linear fit in the scattering region, see Fig~\ref{fig:TGrad} orange line. The established gradient is very close to $\frac{T_{\rm H}-T_{\rm C}}{L}$, where $L$ is the distance between the leads, and $T_{\rm H}=525$ and $T_{\rm C}=475$~K are the target temperatures of the leads. These temperatures keep all parts of the simulation cell in the cubic crystal phase. By applying weakly coupled Langevin thermostats to the leads, a temperature profile without interface effects \cite{Dunn:jap16,Li:jcp19,Zhou:prb19} or interface thermal resistances at the lead$|$scattering-region interface is obtained, see SI. 

The computational efficiency of the MLFF method allows the simulation of atomic scale heat flow in large $4\times 4\times 20$ supercells of the cubic unit cell. To converge to the bulk conductivity value $L$ should be (much) larger than the phonon mean free path ($\Lambda$). In this work, we assume that the length of the scattering region ($L=8a_{\rm lat}\approx 46\AA$) is sufficient, considering the already very low $\Lambda$ values reported for CsPbBr$_3$ at 350~K in Ref.~\cite{Simoncelli:prx22}. See the SI for a comparison to $L=3a_{\rm lat}$ simulations, which result in comparable $\kappa$ values.\newline

\noindent The obtained thermal conductivity values for the different independent simulations are tabulated in Table~\ref{tbl:1}. We determine the value of the thermal conductivity $\kappa$ for CsPbBr$_3$ by the mean of the five independent simulations to be $0.53\frac{\rm W}{\rm m K}$, with a 95\% uncertainty interval of $\pm0.044$.  This is in agreement with experimental values~\cite{Lee:pnas17,Wang:nanol18,Kanak:aip20} and slightly above the modeled thermal conductivity coefficient of Simoncelli~\textit{et al.}~\cite{Simoncelli:natp19} (who reported a value of $\sim{}0.4\frac{\rm W}{\rm m K}$) and within the spread of the three approaches by Tadano~\textit{et.al.}~\cite{Tadano:prl22}. Our simulations indicate that decreasing the mass of the Cs atoms (blue shifting $\omega_{\rm E}$)  by an order of magnitude only changes the thermal conductivity slightly, resulting in $0.58\pm0.049 \frac{\rm W}{\rm m K}$. This is a small but statistically significant increase in $\kappa$. Our simulations also show that increasing the mass of the Cs atoms (red shifting $\omega_{\rm E}$)  by an order of magnitude changes the thermal conductivity more substantially, resulting in $\sim0.36 \frac{\rm W}{\rm m K}$. These observations combined show that in this crystal that exhibits ultra-low $\kappa$, merely tuning the Einstein rattling frequency still has a noticeable effect on the thermal conductivity.
\newline

 We will now analyze whether the high/low mass system enhances/decreases scattering with low frequency acoustic modes, respectively. From a Peierls-Boltzmann (PB)-transport point of view this would either decrease or enhance $\kappa$. We will first establish whether the acoustic modes are affected, by comparing the acoustic branches of the Pb VACF spectra in Figs. \ref{fig:CsPb}(d-f) close to $\Gamma$, we observe a reduced intensity of the acoustic bands in Fig.~\ref{fig:CsPb}(f). This perturbation of the acoustic modes occurs slightly off the $\Gamma$ point precisely around $\omega_{\rm E}=\sim0.3$~THz, whereas the higher frequencies are largely unchanged. A similar effect is observed in the Br spectra, see SI. The coupling of the heavy mass rattling band with the acoustic mode can be more clearly identified in Figure~\ref{fig:coupling}(a). In the power spectra of the first acoustic phonon branch $\alpha=1$ (black lines), we observe Lorentzian line-shapes for both the light and the normal Cs mass along the whole $\Gamma$-X line in $\mathbf{q}$-space. This indicates that $\alpha=1$ is a well-defined phonon-quasi-particle, albeit increasingly short lived as indicated by their lifetimes shown in Fig.~\ref{fig:coupling}(b). For the heavy mass system we observe multi-peak spectra at $\mathbf{q}=[\frac{1}{5} 0 0]$ and $[\frac{3}{10} 0 0]$. Around $\sim0.7$~THz a second peak appears that is much wider. This indicates that the acoustic mode is coupled with low lying optical modes, which makes the Lorentzian lineshape approximation not applicable here  and it cannot be uniquely fitted. Spectra of the other $\alpha=2,3$ acoustic branches are in the SI and show similar behavior. Having established this, we fit Lorentzians (colored lines) to the highest peak in each of the 15 spectra. We extract all required parameters for the PB conductivity of Eq.~(\ref{eq:P}): the central frequency $\omega$ of the peak, the occupancy $n$ from the peak height, and the phonon lifetime $\tau$ from the full width at half maximum $(\Gamma)$ as $\tau=\frac{1}{2\Gamma}$. The resulting frequencies and lifetimes are shown in Fig.~\ref{fig:coupling}(b)  and the relative peak heights can be read off Fig.~\ref{fig:coupling}(a), taking into account the indicated magnification factors. The dispersion of the acoustic mode as observed in the normal and light mass system are identical and lie below the rattling frequency $\omega_{\rm E}$ (blue dashed line) of the normal mass system. However, $\omega_{\rm E}$ of the heavy mass system (green dashed line) intersects with the acoustic band. As a result, the dispersion splits, showing a red shift close to $\Gamma$, which turns into a blue shift upon approaching X. The lifetimes of the acoustic phonons rapidly decrease with the distance away from $\mathbf{q}=\Gamma$. 

\begin{figure}[t!]
 \centering
 \includegraphics[width=\columnwidth]{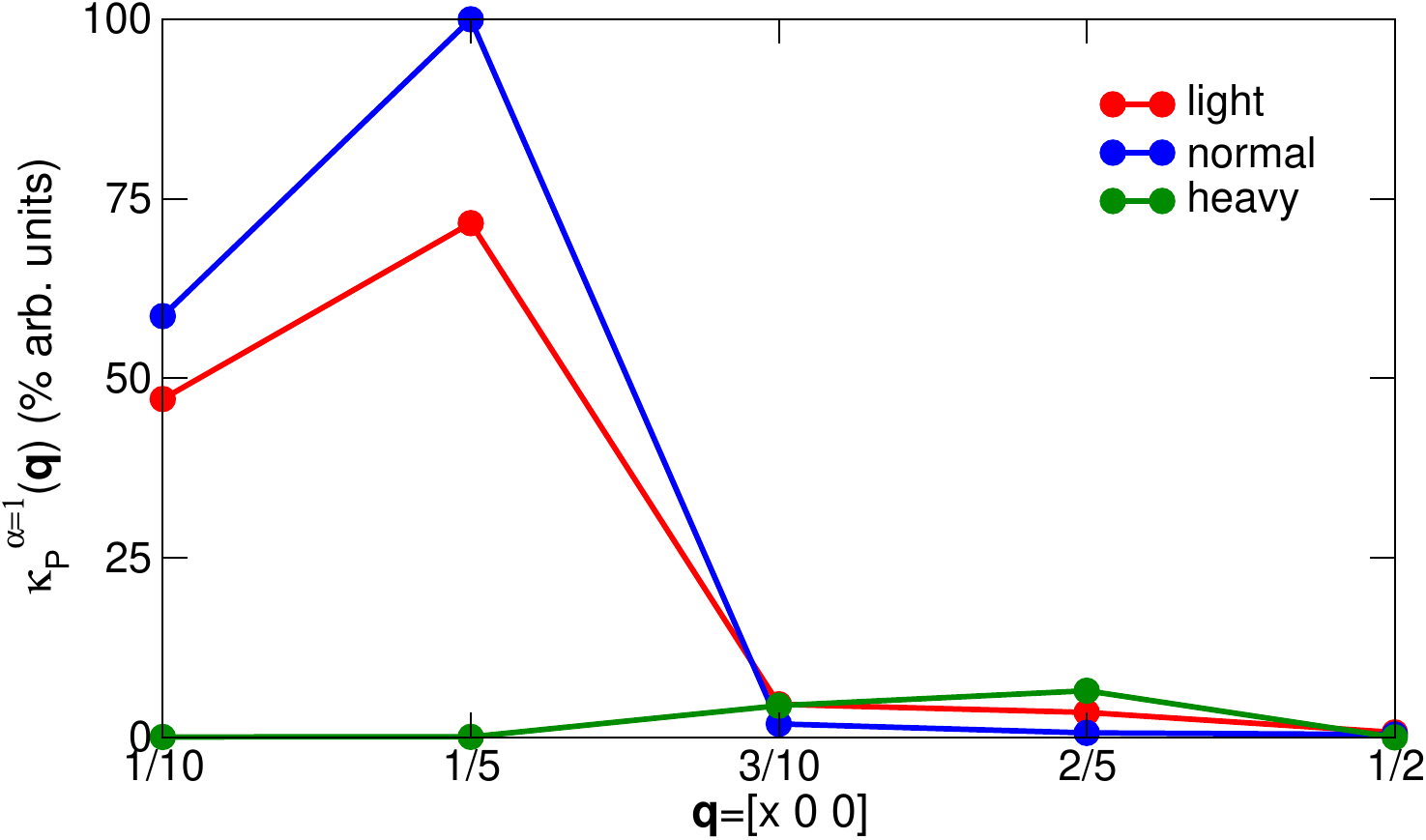}
\caption{Peierls-Boltzmann thermal conductivity based on the Lorentzian fits of PVACFs of the first acoustic mode ($\alpha=1$).} 
\label{fig:peierls}
\end{figure}

These results are combined in the calculated PB conductivity of this mode $\kappa_{\rm P}^{\alpha{}=1}(\mathbf{q})$ as shown in Figure~\ref{fig:peierls}. Details of the computation have been added to the SI. The PB conductivity at low $\mathbf{q}$ is significantly reduced for the heavy mass compared to the normal mass system, whereas it is of the same order when approaching the X-point. This is mainly due to the mode's low occupancy and further reduced by the lower group velocity. The $\kappa_{\rm P}^{\alpha{}=1}$ of the low mass system shows comparable behavior to the normal mass system.

\vspace{2mm}
\section{Discussion}
The ML MD approach applied here presents an ultra-low lattice thermal conductivity of CsPbBr$_3$ at 500~K that is slightly higher than the available experimental values at room temperature. This is expected, since the simulation covers a single crystalline domain without impurities, defects and grain boundaries. The presented lifetimes and lattice thermal conductivities are of the same order as those in recent lattice thermal conductivity model works of Refs.~\cite{Simoncelli:prx22,Tadano:prl22,Wang:prb23}. The model of Simoncelli~\textit{et.al.} shows that only $\sim30\%$ of the thermal conductivity of orthorhombic CsPbBr$_3$ at 300~K originates from $\kappa_{\rm P}$, and most of it is linked to the acoustic modes\cite{Simoncelli:natp19}. The remaining $\sim70\%$ originates from $\kappa_{\rm C}$. The model of Wang~\textit{et.al.} supports this assessment, stating that in cubic CsPbBr$_3$ optical phonons, ranging from 3.1 to 3.9~THz, dominate the heat transport\cite{Wang:prb23}. On the contrary, Tadano~\textit{et.al.} report that $\sim90\%$ of the thermal conductivity of cubic CsPbBr$_3$ at 500~K originates from $\kappa_{\rm P}$, dominated by modes below $\sim 1.5$~THz\cite{Tadano:prl22}. It should be noted that these works apply different levels of theory to obtain the phonon basis.

The coherences' contribution $\kappa_{\rm C}$ was not calculated, since the strongly interacting phonons observed in the MD make a decomposition into projected phonon frequencies and lifetimes far from unique. The fingerprints of these strong interactions are evidenced by the non-Lorentzian lineshapes that are found (apart for some low frequency acoustic modes) across most of the spectrum\cite{Lahnsteiner:prb22}. Therefore, with the current  NEMD study we can not make a quantitative statement about the relative importance of $\kappa_{\rm C}$ vs. $\kappa_{\rm P}$. However, the observation made here that the thermal conductivity can still be significantly tuned, and that this can be explained by combining the rattling picture with a PB-view on thermal transport, seems to advocate for a $\kappa_{\rm P}$ contribution of importance.  We note that we cannot rule out a back-coupling effect, where the decrease of $\kappa_{\rm P}$ affects the value of $\kappa_{\rm C}$. We would therefore highly encourage a study with lattice models making the same alterations to the Cs masses as in the present study. In future work, a modal decomposition of the thermal conductance \cite{Saaskilahti:prb14} or conductivity\cite{Lv:njp16} as computed by MD could help in resolving this question.

\vspace{2mm}
\section{Conclusion}
A near first-principles accuracy ML interatomic potential has been used to simulate large thermodynamic ensembles of the cubic phase of CsPbBr$_3$. The Cs cations introduce dynamic disorder into the crystal, and, depending on their mass, rattle at different effective frequencies.  We have shown that the ultra-low lattice thermal conductivity can be tuned by creating a substantial change in the Einstein oscillator (rattling) frequency. The lattice thermal conductivity was computed to be $0.53\pm0.04\frac{\rm W}{\rm m K}$ for normal Cs mass and reduced to $0.36\pm0.02\frac{\rm W}{\rm m K}$ for a Cs mass ten times heavier. By lowering the mass to a tenth of normal Cs we shift the rattling band far up in the spectrum. However, only a slight increase of the thermal conductivity to $0.58\pm0.05\frac{\rm W}{\rm m K}$ is observed, which indicates that the main cause for the ultra-low thermal conductivity is not incoherent scattering with guest Einstein modes. Rather, the high degree of mode coupling across the spectrum results in a phonon liquid in which diffusive heat transport dominates. These results provide new insights applicable to the wider class of halide perovskites. For practical thermoelectric engineering purposes we have shown that reducing $\kappa$ by a suitably chosen cation based on its rattling frequency in the Pb-Br framework  is possible, but has only a limited effect. However, this does not exclude chemical possibilities, such as cation replacement Cs$\rightarrow X$, whereby $X$ is non-spherical or has different volume/internal structure\cite{Yuping:com14,Caddeo:pccp16,Bokdam:jpcc21}, introducing multiple species $X'$, and mixing the metal and halide compositions. 

\section{Data Availability Statement}
Research data for this paper has been made available through a data set in the 4TU.ResearchData repository; see Ref.~\cite{Bokdam:4TUDATA23}. The following data is stored: (i) The electronic structure databases, including crystal structures and corresponding DFT energies, forces, and stress tensors, used to train the MLFFs (ML\_AB files). (ii) A high level analysis of the electronic structure databases presented by pdf fact sheets generated with open-source \textsc{fpdataviewer} software\cite{FPdataViewer23}. (iii) The generated ML force fields (ML\_FF files) applied in the NEMD simulations in this study.

\section{Supporting Information}
Computational details of the MLFF training procedure, an error analysis of the generated MLFFs, the contributions of Cs, Pb and Br to the $\mathbf{q}$-VACF spectra, results of test calculation to determine the computational setup of the active measurements, additional NEMD simulation results for a small $4\times4\times10$ supercell, explanation of the applied Lorentzian fitting procedure of the PVACF spectra, additional PVACF spectra for the three acoustic modes, and an explanation of the Peierls’ conductivity calculation.

\begin{acknowledgement}
 This work was sponsored by NWO Domain Science for the use of supercomputer facilities. 
 \end{acknowledgement}
\providecommand{\latin}[1]{#1}
\makeatletter
\providecommand{\doi}
  {\begingroup\let\do\@makeother\dospecials
  \catcode`\{=1 \catcode`\}=2 \doi@aux}
\providecommand{\doi@aux}[1]{\endgroup\texttt{#1}}
\makeatother
\providecommand*\mcitethebibliography{\thebibliography}
\csname @ifundefined\endcsname{endmcitethebibliography}
  {\let\endmcitethebibliography\endthebibliography}{}


\begin{mcitethebibliography}{72}
\providecommand*\natexlab[1]{#1}
\providecommand*\mciteSetBstSublistMode[1]{}
\providecommand*\mciteSetBstMaxWidthForm[2]{}
\providecommand*\mciteBstWouldAddEndPuncttrue
  {\def\EndOfBibitem{\unskip.}}
\providecommand*\mciteBstWouldAddEndPunctfalse
  {\let\EndOfBibitem\relax}
\providecommand*\mciteSetBstMidEndSepPunct[3]{}
\providecommand*\mciteSetBstSublistLabelBeginEnd[3]{}
\providecommand*\EndOfBibitem{}
\mciteSetBstSublistMode{f}
\mciteSetBstMaxWidthForm{subitem}{(\alph{mcitesubitemcount})}
\mciteSetBstSublistLabelBeginEnd
  {\mcitemaxwidthsubitemform\space}
  {\relax}
  {\relax}

\bibitem[Simoncelli \latin{et~al.}(2019)Simoncelli, Marzari, and Mauri]{Simoncelli:natp19} Simoncelli,~M.; Marzari,~N.; Mauri,~F. Unified theory of thermal transport in crystals and glasses. \emph{Nature Phys.} \textbf{2019}, \emph{15}, 809--813\relax
\mciteBstWouldAddEndPuncttrue
\mciteSetBstMidEndSepPunct{\mcitedefaultmidpunct}
{\mcitedefaultendpunct}{\mcitedefaultseppunct}\relax
\EndOfBibitem
\bibitem[Miyata \latin{et~al.}(2017)Miyata, Atallah, and Zhu]{Miyata:scadv17} Miyata,~K.; Atallah,~T.~L.; Zhu,~X.-Y. Lead halide perovskites: Crystal-liquid duality, phonon glass electron crystals, and large polaron formation. \emph{Sci. Adv.} \textbf{2017}, \emph{3}, e1701469\relax
\mciteBstWouldAddEndPuncttrue
\mciteSetBstMidEndSepPunct{\mcitedefaultmidpunct}
{\mcitedefaultendpunct}{\mcitedefaultseppunct}\relax
\EndOfBibitem
\bibitem[Lanigan-Atkins \latin{et~al.}(2021)Lanigan-Atkins, He, Krogstad, Pajerowski, Abernathy, Xu, Xu, Chung, Kanatzidis, Rosenkranz, Osborn, and Delaire]{Lanigan-Atkins:natm21} Lanigan-Atkins,~T.; He,~X.; Krogstad,~M.~J.; Pajerowski,~D.~M.; Abernathy,~D.~L.; Xu,~G. N. M.~N.; Xu,~Z.; Chung,~D.-Y.; Kanatzidis,~M.~G.; Rosenkranz,~S. \latin{et~al.}  Two-dimensional overdamped fluctuations of the soft perovskite lattice in CsPbBr3. \emph{Nature Mater.} \textbf{2021}, \emph{20}, 977--983\relax
\mciteBstWouldAddEndPuncttrue
\mciteSetBstMidEndSepPunct{\mcitedefaultmidpunct}
{\mcitedefaultendpunct}{\mcitedefaultseppunct}\relax
\EndOfBibitem
\bibitem[Gehrmann and Egger(2019)Gehrmann, and Egger]{Gehrmann:natc:19}
Gehrmann,~C.; Egger,~D.~A. Dynamic shortening of disorder potentials in anharmonic halide perovskites. \emph{Nature Comm.} \textbf{2019}, \emph{10}, 3141\relax
\mciteBstWouldAddEndPuncttrue
\mciteSetBstMidEndSepPunct{\mcitedefaultmidpunct}
{\mcitedefaultendpunct}{\mcitedefaultseppunct}\relax
\EndOfBibitem
\bibitem[Songvilay \latin{et~al.}(2019)Songvilay, Giles-Donovan, Bari, Ye, Minns, Green, Xu, Gehring, Schmalzl, Ratcliff, Brown, Chernyshov, Beek, Cochran, and Stock]{Songvilay:prm19} Songvilay,~M.; Giles-Donovan,~N.; Bari,~M.; Ye,~G.,~Z.; Minns,~L.,~J.; Green,~M.; Xu,~G.; Gehring,~M.,~P.; Schmalzl,~K.; Ratcliff,~D.,~W. \latin{et~al.}  Common acoustic phonon lifetimes in inorganic and hybrid lead halide perovskites. \emph{Phys. Rev. Mat.} \textbf{2019}, \emph{3}, 093602\relax
\mciteBstWouldAddEndPuncttrue
\mciteSetBstMidEndSepPunct{\mcitedefaultmidpunct}
{\mcitedefaultendpunct}{\mcitedefaultseppunct}\relax
\EndOfBibitem
\bibitem[Jinnouchi \latin{et~al.}(2019)Jinnouchi, Lahnsteiner, Karsai, Kresse, and Bokdam]{Jinnouchi:prl19} Jinnouchi,~R.; Lahnsteiner,~J.; Karsai,~F.; Kresse,~G.; Bokdam,~M. Phase Transitions of Hybrid Perovskites Simulated by Machine-Learning Force Fields Trained on the Fly with Bayesian Inference. \emph{Phys. Rev. Lett.} \textbf{2019}, \emph{122}, 225701\relax
\mciteBstWouldAddEndPuncttrue
\mciteSetBstMidEndSepPunct{\mcitedefaultmidpunct}
{\mcitedefaultendpunct}{\mcitedefaultseppunct}\relax
\EndOfBibitem
\bibitem[Zhu \latin{et~al.}(2022)Zhu, Caicedo-Dávila, Gehrmann, and Egger]{Zhu:acsami22} Zhu,~X.; Caicedo-Dávila,~S.; Gehrmann,~C.; Egger,~D.~A. Probing the Disorder Inside the Cubic Unit Cell of Halide Perovskites from First-Principles. \emph{ACS Appl. Mater. Interfaces} \textbf{2022}, \emph{14}, 22973--22981\relax
\mciteBstWouldAddEndPuncttrue
\mciteSetBstMidEndSepPunct{\mcitedefaultmidpunct}
{\mcitedefaultendpunct}{\mcitedefaultseppunct}\relax
\EndOfBibitem
\bibitem[Eaton \latin{et~al.}(2016)Eaton, Lai, Gibson, Wong, Dou, Ma, Wang, Leone, and Yang]{Eaton:pnas16} Eaton,~S.~W.; Lai,~M.; Gibson,~N.~A.; Wong,~A.~B.; Dou,~L.; Ma,~J.; Wang,~L.-W.; Leone,~S.~R.; Yang,~P. Lasing in robust cesium lead halide perovskite nanowires. \emph{Proc. Natl. Acad. Sci. U.S.A.} \textbf{2016}, \emph{113}, 1993--1998\relax
\mciteBstWouldAddEndPuncttrue
\mciteSetBstMidEndSepPunct{\mcitedefaultmidpunct}
{\mcitedefaultendpunct}{\mcitedefaultseppunct}\relax
\EndOfBibitem
\bibitem[Chen \latin{et~al.}(2017)Chen, Fu, Samad, Dang, Zhao, Shen, Guo, and Jin]{Chen:nanol17} Chen,~J.; Fu,~Y.; Samad,~L.; Dang,~L.; Zhao,~Y.; Shen,~S.; Guo,~L.; Jin,~S. Vapor-Phase Epitaxial Growth of Aligned Nanowire Networks of Cesium Lead Halide Perovskites (CsPbX3, X = Cl, Br, I). \emph{Nano Lett.} \textbf{2017}, \emph{17}, 460--466\relax
\mciteBstWouldAddEndPuncttrue
\mciteSetBstMidEndSepPunct{\mcitedefaultmidpunct}
{\mcitedefaultendpunct}{\mcitedefaultseppunct}\relax
\EndOfBibitem
\bibitem[Wang \latin{et~al.}(2018)Wang, Lin, Zhu, Zheng, Wang, Li, and Zhu]{Wang:nanol18} Wang,~Y.; Lin,~R.; Zhu,~P.; Zheng,~Q.; Wang,~Q.; Li,~D.; Zhu,~J. Cation Dynamics Governed Thermal Properties of Lead Halide Perovskite Nanowires. \emph{Nano Lett.} \textbf{2018}, \emph{18}, 2772--2779\relax
\mciteBstWouldAddEndPuncttrue
\mciteSetBstMidEndSepPunct{\mcitedefaultmidpunct}
{\mcitedefaultendpunct}{\mcitedefaultseppunct}\relax
\EndOfBibitem
\bibitem[Yuping and Giulia(2014)Yuping, and Giulia]{Yuping:com14} Yuping,~H.; Giulia,~G. Perovskites for Solar Thermoelectric Applications: A First Principle Study of CH3NH3AI3 (A = Pb and Sn). \emph{Chem. Mater.} \textbf{2014}, \emph{26}, 5394\relax
\mciteBstWouldAddEndPuncttrue
\mciteSetBstMidEndSepPunct{\mcitedefaultmidpunct}
{\mcitedefaultendpunct}{\mcitedefaultseppunct}\relax
\EndOfBibitem
\bibitem[Mettan \latin{et~al.}(2015)Mettan, Pisoni, Jacimovic, Nafradi, Spina, Pavuna, Forro, and Horvath]{Mettan:jpcc15} Mettan,~X.; Pisoni,~A.; Jacimovic,~J.; Nafradi,~B.; Spina,~M.; Pavuna,~D.; Forro,~L.; Horvath,~E. Tuning of the Thermoelectric Figure of Merit of CH3NH3MI3 (M=Pb,Sn) Photovoltaic Perovskites. \emph{J. Phys. Chem. C} \textbf{2015}, \emph{119}, 11506\relax
\mciteBstWouldAddEndPuncttrue
\mciteSetBstMidEndSepPunct{\mcitedefaultmidpunct}
{\mcitedefaultendpunct}{\mcitedefaultseppunct}\relax
\EndOfBibitem
\bibitem[Yue \latin{et~al.}(2016)Yue, Zhang, Qin, Yang, and Hu]{Yue:prb16} Yue,~S.-Y.; Zhang,~X.; Qin,~G.; Yang,~J.; Hu,~M. Insight into the collective vibrational modes driving ultralow thermal conductivity of perovskite solar cells. \emph{Phys. Rev. B} \textbf{2016}, \emph{94}, 115427\relax
\mciteBstWouldAddEndPuncttrue
\mciteSetBstMidEndSepPunct{\mcitedefaultmidpunct}
{\mcitedefaultendpunct}{\mcitedefaultseppunct}\relax
\EndOfBibitem
\bibitem[Filippetti \latin{et~al.}(2016)Filippetti, Caddeo, Delugas, and Mattoni]{Filippetti:jpcc16} Filippetti,~A.; Caddeo,~C.; Delugas,~P.; Mattoni,~A. Appealing Perspectives of Hybrid Lead–Iodide Perovskites as Thermoelectric Materials. \emph{J. Phys. Chem. C} \textbf{2016}, \emph{120}, 28472\relax
\mciteBstWouldAddEndPuncttrue
\mciteSetBstMidEndSepPunct{\mcitedefaultmidpunct}
{\mcitedefaultendpunct}{\mcitedefaultseppunct}\relax
\EndOfBibitem
\bibitem[Guo \latin{et~al.}(2017)Guo, Xia, Gong, Stoumpos, McCall, Alexander, Ma, Zhou, Gosztola, Ketterson, Kanatzidis, Xu, Chan, and Schaller]{Guo:acsel17} Guo,~P.; Xia,~Y.; Gong,~J.; Stoumpos,~C.~C.; McCall,~K.~M.; Alexander,~G. C.~B.; Ma,~Z.; Zhou,~H.; Gosztola,~D.~J.; Ketterson,~J.~B. \latin{et~al.} Polar Fluctuations in Metal Halide Perovskites Uncovered by Acoustic Phonon Anomalies. \emph{ACS Energy Lett.} \textbf{2017}, \emph{2}, 2463--2469\relax
\mciteBstWouldAddEndPuncttrue
\mciteSetBstMidEndSepPunct{\mcitedefaultmidpunct}
{\mcitedefaultendpunct}{\mcitedefaultseppunct}\relax
\EndOfBibitem
\bibitem[Lee \latin{et~al.}(2017)Lee, Li, Wong, Zhang, Lai, Yu, Kong, Lin, Urban, Grossman, and Yang]{Lee:pnas17} Lee,~W.; Li,~H.; Wong,~A.~B.; Zhang,~D.; Lai,~M.; Yu,~Y.; Kong,~Q.; Lin,~E.; Urban,~J.~J.; Grossman,~J.~C. \latin{et~al.}  Ultralow thermal conductivity in all-inorganic halide perovskites. \emph{Proc. Natl. Acad. Sci. U.S.A.} \textbf{2017}, \emph{114}, 8693--8697\relax
\mciteBstWouldAddEndPuncttrue
\mciteSetBstMidEndSepPunct{\mcitedefaultmidpunct}
{\mcitedefaultendpunct}{\mcitedefaultseppunct}\relax
\EndOfBibitem
\bibitem[Xing \latin{et~al.}(2013)Xing, Mathews, Sun, Lim, Lam, Gr{\"a}tzel, Mhaisalkar, and Sum]{Xing:sc13} Xing,~G.; Mathews,~N.; Sun,~S.; Lim,~S.~S.; Lam,~Y.~M.; Gr{\"a}tzel,~M.; Mhaisalkar,~S.; Sum,~T.~C. Long-Range Balanced Electron- and Hole-Transport Lengths in Organic-Inorganic CH3NH3PbI3. \emph{Science} \textbf{2013}, \emph{342}, 344--347\relax
\mciteBstWouldAddEndPuncttrue
\mciteSetBstMidEndSepPunct{\mcitedefaultmidpunct}
{\mcitedefaultendpunct}{\mcitedefaultseppunct}\relax
\EndOfBibitem
\bibitem[Xie \latin{et~al.}(2020)Xie, Hao, Bao, Slade, Snyder, Wolverton, and Kanatzidis]{Xie:jacs20} Xie,~H.; Hao,~S.; Bao,~J.; Slade,~T.~J.; Snyder,~G.~J.; Wolverton,~C.; Kanatzidis,~M.~G. All-Inorganic Halide Perovskites as Potential Thermoelectric Materials: Dynamic Cation off-Centering Induces Ultralow Thermal Conductivity. \emph{J. Am. Chem. Soc.} \textbf{2020}, \emph{142}, 9553--9563\relax
\mciteBstWouldAddEndPuncttrue
\mciteSetBstMidEndSepPunct{\mcitedefaultmidpunct}
{\mcitedefaultendpunct}{\mcitedefaultseppunct}\relax
\EndOfBibitem
\bibitem[Jong \latin{et~al.}(2022)Jong, Kim, Ri, Kye, Pak, Cottenier, and Yu]{Jong:chc22} Jong,~U.-G.; Kim,~Y.-S.; Ri,~C.-H.; Kye,~Y.-H.; Pak,~C.-J.; Cottenier,~S.; Yu,~C.-J. Twofold rattling mode-induced ultralow thermal conductivity in vacancy-ordered double perovskite Cs2SnI6. \emph{Chem. Commun.} \textbf{2022}, \emph{58}, 4223--4226\relax
\mciteBstWouldAddEndPuncttrue
\mciteSetBstMidEndSepPunct{\mcitedefaultmidpunct}
{\mcitedefaultendpunct}{\mcitedefaultseppunct}\relax
\EndOfBibitem
\bibitem[Acharyya \latin{et~al.}(2022)Acharyya, Ghosh, Pal, Rana, Dutta, Swain, Etter, Soni, Waghmare, and Biswas]{Acharyya:natc22} Acharyya,~P.; Ghosh,~T.; Pal,~K.; Rana,~K.~S.; Dutta,~M.; Swain,~D.; Etter,~M.; Soni,~A.; Waghmare,~U.~V.; Biswas,~K. Glassy thermal conductivity in Cs3Bi2I6Cl3 single crystal. \emph{Nature Comm.} \textbf{2022}, \emph{13}, 5053\relax
\mciteBstWouldAddEndPuncttrue
\mciteSetBstMidEndSepPunct{\mcitedefaultmidpunct}
{\mcitedefaultendpunct}{\mcitedefaultseppunct}\relax
\EndOfBibitem
\bibitem[Hermann \latin{et~al.}(2003)Hermann, Jin, Schweika, Grandjean, Mandrus, Sales, and Long]{Hermann:prl03} Hermann,~R.~P.; Jin,~R.; Schweika,~W.; Grandjean,~F.; Mandrus,~D.; Sales,~B.~C.; Long,~G.~J. Einstein Oscillators in Thallium Filled Antimony Skutterudites. \emph{Phys. Rev. Lett.} \textbf{2003}, \emph{90}, 135505\relax
\mciteBstWouldAddEndPuncttrue
\mciteSetBstMidEndSepPunct{\mcitedefaultmidpunct}
{\mcitedefaultendpunct}{\mcitedefaultseppunct}\relax
\EndOfBibitem
\bibitem[Ge \latin{et~al.}(2022)Ge, Li, Feng, Zheng, Jia, Wu, and Jin]{Ge:aem22} Ge,~Z.-H.; Li,~W.-J.; Feng,~J.; Zheng,~F.; Jia,~C.-L.; Wu,~D.; Jin,~L. Atomic-Scale Observation of Off-Centering Rattlers in Filled Skutterudites. \emph{Adv. Energy Mater.} \textbf{2022}, \emph{12}, 2103770\relax
\mciteBstWouldAddEndPuncttrue
\mciteSetBstMidEndSepPunct{\mcitedefaultmidpunct}
{\mcitedefaultendpunct}{\mcitedefaultseppunct}\relax
\EndOfBibitem
\bibitem[Chen \latin{et~al.}(2018)Chen, Zhang, and Chen]{Chen:fier18} Chen,~C.; Zhang,~Z.; Chen,~J. Revisit to the Impacts of Rattlers on Thermal Conductivity of Clathrates. \emph{Front. Energy Res.} \textbf{2018}, \emph{6}\relax
\mciteBstWouldAddEndPuncttrue
\mciteSetBstMidEndSepPunct{\mcitedefaultmidpunct}
{\mcitedefaultendpunct}{\mcitedefaultseppunct}\relax
\EndOfBibitem
\bibitem[Takabatake \latin{et~al.}(2014)Takabatake, Suekuni, Nakayama, and Kaneshita]{Takabatake:rmp14} Takabatake,~T.; Suekuni,~K.; Nakayama,~T.; Kaneshita,~E. Phonon-glass electron-crystal thermoelectric clathrates: Experiments and theory. \emph{Rev. Mod. Phys.} \textbf{2014}, \emph{86}, 669--716\relax
\mciteBstWouldAddEndPuncttrue
\mciteSetBstMidEndSepPunct{\mcitedefaultmidpunct}
{\mcitedefaultendpunct}{\mcitedefaultseppunct}\relax
\EndOfBibitem
\bibitem[Hermann \latin{et~al.}(2004)Hermann, Grandjean, and Long]{Hermann:ajp04} Hermann,~P.,~R.; Grandjean,~F.; Long,~J.,~G. Einstein oscillators that impede thermal transport. \emph{Am. J. Phys.} \textbf{2004}, \emph{73}, 110--118\relax
\mciteBstWouldAddEndPuncttrue
\mciteSetBstMidEndSepPunct{\mcitedefaultmidpunct}
{\mcitedefaultendpunct}{\mcitedefaultseppunct}\relax
\EndOfBibitem
\bibitem[Dong \latin{et~al.}(2001)Dong, Sankey, and Myles]{Dong:prl01} Dong,~J.; Sankey,~O.~F.; Myles,~C.~W. Theoretical Study of the Lattice Thermal Conductivity in Ge Framework Semiconductors. \emph{Phys. Rev. Lett.} \textbf{2001}, \emph{86}, 2361--2364\relax
\mciteBstWouldAddEndPuncttrue
\mciteSetBstMidEndSepPunct{\mcitedefaultmidpunct}
{\mcitedefaultendpunct}{\mcitedefaultseppunct}\relax
\EndOfBibitem
\bibitem[Koza \latin{et~al.}(2008)Koza, Johnson, Viennois, Mutka, Girard, and Ravot]{Koza:natm08} Koza,~M.~M.; Johnson,~M.~R.; Viennois,~R.; Mutka,~H.; Girard,~L.; Ravot,~D. Breakdown of phonon glass paradigm in La- and Ce-filled Fe4Sb12 skutterudites. \emph{Nature Mater.} \textbf{2008}, \emph{7}, 805--810\relax
\mciteBstWouldAddEndPuncttrue
\mciteSetBstMidEndSepPunct{\mcitedefaultmidpunct}
{\mcitedefaultendpunct}{\mcitedefaultseppunct}\relax
\EndOfBibitem
\bibitem[Christensen \latin{et~al.}(2008)Christensen, Abrahamsen, Christensen, Juranyi, Andersen, Lefmann, Andreasson, Bahl, and Iversen]{Christensen:natm08} Christensen,~M.; Abrahamsen,~A.~B.; Christensen,~N.~B.; Juranyi,~F.; Andersen,~N.~H.; Lefmann,~K.; Andreasson,~J.; Bahl,~C. R.~H.; Iversen,~B.~B. Avoided crossing of rattler modes in thermoelectric materials. \emph{Nature Mater.} \textbf{2008}, \emph{7}, 811--815\relax
\mciteBstWouldAddEndPuncttrue
\mciteSetBstMidEndSepPunct{\mcitedefaultmidpunct}
{\mcitedefaultendpunct}{\mcitedefaultseppunct}\relax
\EndOfBibitem
\bibitem[Toberer \latin{et~al.}(2011)Toberer, Zevalkink, and Snyder]{Toberer:jmc11} Toberer,~E.~S.; Zevalkink,~A.; Snyder,~G.~J. Phonon engineering through crystal chemistry. \emph{J. Mater. Chem.} \textbf{2011}, \emph{21}, 15843--15852\relax
\mciteBstWouldAddEndPuncttrue
\mciteSetBstMidEndSepPunct{\mcitedefaultmidpunct}
{\mcitedefaultendpunct}{\mcitedefaultseppunct}\relax
\EndOfBibitem
\bibitem[Li and Mingo(2014)Li, and Mingo]{Wu:prb14} Li,~W.; Mingo,~N. Thermal conductivity of fully filled skutterudites: Role of the filler. \emph{Phys. Rev. B} \textbf{2014}, \emph{89}, 184304\relax
\mciteBstWouldAddEndPuncttrue
\mciteSetBstMidEndSepPunct{\mcitedefaultmidpunct}
{\mcitedefaultendpunct}{\mcitedefaultseppunct}\relax
\EndOfBibitem
\bibitem[Tadano \latin{et~al.}(2015)Tadano, Gohda, and Tsuneyuki]{Tadano:prb15} Tadano,~T.; Gohda,~Y.; Tsuneyuki,~S. Impact of Rattlers on Thermal Conductivity of a Thermoelectric Clathrate: A First-Principles Study. \emph{Phys. Rev. Lett.} \textbf{2015}, \emph{114}, 095501\relax
\mciteBstWouldAddEndPuncttrue
\mciteSetBstMidEndSepPunct{\mcitedefaultmidpunct}
{\mcitedefaultendpunct}{\mcitedefaultseppunct}\relax
\EndOfBibitem
\bibitem[Lahnsteiner and Bokdam(2022)Lahnsteiner, and Bokdam]{Lahnsteiner:prb22} Lahnsteiner,~J.; Bokdam,~M. Anharmonic lattice dynamics in large thermodynamic ensembles with machine-learning force fields: $\mathrm{Cs}\mathrm{Pb}{\mathrm{Br}}_{3}$, a phonon liquid with Cs rattlers. \emph{Phys. Rev. B} \textbf{2022}, \emph{105}, 024302\relax
\mciteBstWouldAddEndPuncttrue
\mciteSetBstMidEndSepPunct{\mcitedefaultmidpunct}
{\mcitedefaultendpunct}{\mcitedefaultseppunct}\relax
\EndOfBibitem
\bibitem[Hellman and Abrikosov(2013)Hellman, and Abrikosov]{Hellman:prb13A} Hellman,~O.; Abrikosov,~I.~A. Temperature-dependent effective third-order interatomic force constants from first principles. \emph{Phys. Rev. B} \textbf{2013}, \emph{88}, 144301\relax
\mciteBstWouldAddEndPuncttrue
\mciteSetBstMidEndSepPunct{\mcitedefaultmidpunct}
{\mcitedefaultendpunct}{\mcitedefaultseppunct}\relax
\EndOfBibitem
\bibitem[Reissland(1973)]{Reissland:Book1973} Reissland,~A.,~J. \emph{The Physics of Phonons}; John Wiley and Sons Ltd., 1973\relax
\mciteBstWouldAddEndPuncttrue
\mciteSetBstMidEndSepPunct{\mcitedefaultmidpunct}
{\mcitedefaultendpunct}{\mcitedefaultseppunct}\relax
\EndOfBibitem
\bibitem[Sun \latin{et~al.}(2010)Sun, Shen, and Allen]{Allen:prb10} Sun,~T.; Shen,~X.; Allen,~P.~B. Phonon quasiparticles and and anharmonic perturbation theory tested by molecular dynamics on a model system. \emph{Phys. Rev. B} \textbf{2010}, \emph{82}, 224304\relax
\mciteBstWouldAddEndPuncttrue
\mciteSetBstMidEndSepPunct{\mcitedefaultmidpunct}
{\mcitedefaultendpunct}{\mcitedefaultseppunct}\relax
\EndOfBibitem
\bibitem[Cowley(1963)]{Cowley:ap63} Cowley,~R.~A. The lattice Dynamics of an Anharmonic Crystal. \emph{Adv. Phys.} \textbf{1963}, \emph{12}, 421--480\relax
\mciteBstWouldAddEndPuncttrue
\mciteSetBstMidEndSepPunct{\mcitedefaultmidpunct}
{\mcitedefaultendpunct}{\mcitedefaultseppunct}\relax
\EndOfBibitem
\bibitem[Tadano and Saidi(2022)Tadano, and Saidi]{Tadano:prl22} Tadano,~T.; Saidi,~W.~A. First-Principles Phonon Quasiparticle Theory Applied to a Strongly Anharmonic Halide Perovskite. \emph{Phys. Rev. Lett.} \textbf{2022}, \emph{129}, 185901\relax
\mciteBstWouldAddEndPuncttrue
\mciteSetBstMidEndSepPunct{\mcitedefaultmidpunct}
{\mcitedefaultendpunct}{\mcitedefaultseppunct}\relax
\EndOfBibitem
\bibitem[Wang \latin{et~al.}(2023)Wang, Gao, Zhu, Ren, Hu, Sun, Ding, Xia, and Li]{Wang:prb23} Wang,~X.; Gao,~Z.; Zhu,~G.; Ren,~J.; Hu,~L.; Sun,~J.; Ding,~X.; Xia,~Y.; Li,~B. Role of high-order anharmonicity and off-diagonal terms in thermal conductivity: A case study of multiphase ${\mathrm{CsPbBr}}_{3}$. \emph{Phys. Rev. B} \textbf{2023}, \emph{107}, 214308\relax
\mciteBstWouldAddEndPuncttrue
\mciteSetBstMidEndSepPunct{\mcitedefaultmidpunct}
{\mcitedefaultendpunct}{\mcitedefaultseppunct}\relax
\EndOfBibitem
\bibitem[Allen \latin{et~al.}(1999)Allen, Feldman, Fabian, and Wooten]{Allen:pmb99} Allen,~P.~B.; Feldman,~J.~L.; Fabian,~J.; Wooten,~F. Diffusons, locons and propagons: Character of atomic vibrations in amorphous Si. \emph{Phil. Mag. B} \textbf{1999}, \emph{79}, 1715--1731\relax
\mciteBstWouldAddEndPuncttrue
\mciteSetBstMidEndSepPunct{\mcitedefaultmidpunct}
{\mcitedefaultendpunct}{\mcitedefaultseppunct}\relax
\EndOfBibitem
\bibitem[Behler and Parrinello(2007)Behler, and Parrinello]{Behler:prl07} Behler,~J.; Parrinello,~M. Generalized Neural-Network Representation of High-Dimensional Potential-Energy Surfaces. \emph{Phys. Rev. Lett.} \textbf{2007}, \emph{98}, 146401\relax
\mciteBstWouldAddEndPuncttrue
\mciteSetBstMidEndSepPunct{\mcitedefaultmidpunct}
{\mcitedefaultendpunct}{\mcitedefaultseppunct}\relax
\EndOfBibitem
\bibitem[Bart\'ok \latin{et~al.}(2010)Bart\'ok, Payne, Kondor, and Cs\'anyi]{Bartok:prl10} Bart\'ok,~A.~P.; Payne,~M.~C.; Kondor,~R.; Cs\'anyi,~G. Gaussian Approximation Potentials: The Accuracy of Quantum Mechanics, without the Electrons. \emph{Phys. Rev. Lett.} \textbf{2010}, \emph{104}, 136403\relax
\mciteBstWouldAddEndPuncttrue
\mciteSetBstMidEndSepPunct{\mcitedefaultmidpunct}
{\mcitedefaultendpunct}{\mcitedefaultseppunct}\relax
\EndOfBibitem
\bibitem[Rupp \latin{et~al.}(2012)Rupp, Tkatchenko, M\"uller, and von Lilienfeld]{Rupp:prl12} Rupp,~M.; Tkatchenko,~A.; M\"uller,~K.-R.; von Lilienfeld,~O.~A. Fast and Accurate Modeling of Molecular Atomization Energies with Machine Learning. \emph{Phys. Rev. Lett.} \textbf{2012}, \emph{108}, 058301\relax
\mciteBstWouldAddEndPuncttrue
\mciteSetBstMidEndSepPunct{\mcitedefaultmidpunct}
{\mcitedefaultendpunct}{\mcitedefaultseppunct}\relax
\EndOfBibitem
\bibitem[Bart\'ok \latin{et~al.}(2013)Bart\'ok, Kondor, and Cs\'anyi]{Bartok:prb13} Bart\'ok,~A.~P.; Kondor,~R.; Cs\'anyi,~G. On representing chemical environments. \emph{Phys. Rev. B} \textbf{2013}, \emph{87}, 184115\relax
\mciteBstWouldAddEndPuncttrue
\mciteSetBstMidEndSepPunct{\mcitedefaultmidpunct}
{\mcitedefaultendpunct}{\mcitedefaultseppunct}\relax
\EndOfBibitem
\bibitem[M\"{u}ller-Plathe(1997)]{Mueller:jcp97} M\"{u}ller-Plathe,~F. A simple nonequilibrium molecular dynamics method for calculating the thermal conductivity. \emph{J. Chem. Phys.} \textbf{1997}, \emph{106}, 6082--6085\relax
\mciteBstWouldAddEndPuncttrue
\mciteSetBstMidEndSepPunct{\mcitedefaultmidpunct}
{\mcitedefaultendpunct}{\mcitedefaultseppunct}\relax
\EndOfBibitem
\bibitem[Hafskjold \latin{et~al.}(1993)Hafskjold, Ikeshoji, and Ratkje]{Hafskjod:mp93} Hafskjold,~B.; Ikeshoji,~T.; Ratkje,~K.,~S. On the molecular mechanism of thermal diffusion in liquids. \emph{Mol. Phys.} \textbf{1993}, \emph{80}, 1389--1412\relax
\mciteBstWouldAddEndPuncttrue
\mciteSetBstMidEndSepPunct{\mcitedefaultmidpunct}
{\mcitedefaultendpunct}{\mcitedefaultseppunct}\relax
\EndOfBibitem
\bibitem[Dunn \latin{et~al.}(2016)Dunn, Antillion, Maassen, Lundstrom, and Strachan]{Dunn:jap16} Dunn,~J.; Antillion,~E.; Maassen,~J.; Lundstrom,~M.; Strachan,~A. Role of energy distribution in contacts on thermal transport in Si: A molecular dynamics study. \emph{J. Appl. Phys.} \textbf{2016}, \emph{120}, 225112\relax
\mciteBstWouldAddEndPuncttrue
\mciteSetBstMidEndSepPunct{\mcitedefaultmidpunct}
{\mcitedefaultendpunct}{\mcitedefaultseppunct}\relax
\EndOfBibitem
\bibitem[Li \latin{et~al.}(2019)Li, Xiong, Sievers, Hu, Fan, Wei, Bao, Chen, Donadio, and Ala-Nissila]{Li:jcp19} Li,~Z.; Xiong,~S.; Sievers,~C.; Hu,~Y.; Fan,~Z.; Wei,~N.; Bao,~H.; Chen,~S.; Donadio,~D.; Ala-Nissila,~T. Influence of thermostatting on nonequilibrium molecular dynamics simulations of heat conduction in solids. \emph{J. Chem. Phys.} \textbf{2019}, \emph{151}, 234105\relax
\mciteBstWouldAddEndPuncttrue
\mciteSetBstMidEndSepPunct{\mcitedefaultmidpunct}
{\mcitedefaultendpunct}{\mcitedefaultseppunct}\relax
\EndOfBibitem
\bibitem[Zhou \latin{et~al.}(2009)Zhou, Aubry, Jones, Greenstein, and Schelling]{Zhou:prb19} Zhou,~X.~W.; Aubry,~S.; Jones,~R.~E.; Greenstein,~A.; Schelling,~P.~K. Towards more accurate molecular dynamics calculation of thermal conductivity: Case study of GaN bulk crystals. \emph{Phys. Rev. B} \textbf{2009}, \emph{79}, 115201\relax
\mciteBstWouldAddEndPuncttrue
\mciteSetBstMidEndSepPunct{\mcitedefaultmidpunct}
{\mcitedefaultendpunct}{\mcitedefaultseppunct}\relax
\EndOfBibitem
\bibitem[Landau and Lifschitz(2016)Landau, and Lifschitz]{Landau:StatPhys1} Landau,~L.; Lifschitz,~E. \emph{Statistische Physik Teil 1}; Europa Lehrmittel, 2016\relax
\mciteBstWouldAddEndPuncttrue
\mciteSetBstMidEndSepPunct{\mcitedefaultmidpunct}
{\mcitedefaultendpunct}{\mcitedefaultseppunct}\relax
\EndOfBibitem
\bibitem[Togo and Tanaka(2015)Togo, and Tanaka]{Togo:phonopy15} Togo,~A.; Tanaka,~I. First principles phonon calculations in materials science. \emph{Scr. Mater.} \textbf{2015}, \emph{108}, 1--5\relax
\mciteBstWouldAddEndPuncttrue
\mciteSetBstMidEndSepPunct{\mcitedefaultmidpunct}
{\mcitedefaultendpunct}{\mcitedefaultseppunct}\relax
\EndOfBibitem
\bibitem[Jinnouchi \latin{et~al.}(2019)Jinnouchi, Karsai, and Kresse]{Jinnouchi:prb19} Jinnouchi,~R.; Karsai,~F.; Kresse,~G. On-the-fly machine learning force field generation: Application to melting points. \emph{Phys. Rev. B} \textbf{2019}, \emph{100}, 014105\relax
\mciteBstWouldAddEndPuncttrue
\mciteSetBstMidEndSepPunct{\mcitedefaultmidpunct}
{\mcitedefaultendpunct}{\mcitedefaultseppunct}\relax
\EndOfBibitem
\bibitem[Jinnouchi \latin{et~al.}(2020)Jinnouchi, Karsai, Verdi, Asahi, and Kresse]{Jinnouchi:jcp20} Jinnouchi,~R.; Karsai,~F.; Verdi,~C.; Asahi,~R.; Kresse,~G. Descriptors representing two- and three-body atomic distributions and their effects on the accuracy of machine-learned inter-atomic potentials. \emph{J. Chem. Phys.} \textbf{2020}, \emph{152}, 234102\relax
\mciteBstWouldAddEndPuncttrue
\mciteSetBstMidEndSepPunct{\mcitedefaultmidpunct}
{\mcitedefaultendpunct}{\mcitedefaultseppunct}\relax
\EndOfBibitem
\bibitem[Kresse and Hafner(1993)Kresse, and Hafner]{vasp-1} Kresse,~G.; Hafner,~J. Ab initio molecular dynamics for liquid metals. \emph{Phys. Rev. B} \textbf{1993}, \emph{47}, 558--561\relax
\mciteBstWouldAddEndPuncttrue
\mciteSetBstMidEndSepPunct{\mcitedefaultmidpunct}
{\mcitedefaultendpunct}{\mcitedefaultseppunct}\relax
\EndOfBibitem
\bibitem[Kresse and Furthm\"{u}ller(1996)Kresse, and Furthm\"{u}ller] {vasp-2} Kresse,~G.; Furthm\"{u}ller,~J. Efficient iterative schemes for ab initio total-energy calculations using a plane-wave basis set. \emph{Phys. Rev. B} \textbf{1996}, \emph{54}, 11169--11186\relax
\mciteBstWouldAddEndPuncttrue
\mciteSetBstMidEndSepPunct{\mcitedefaultmidpunct}
{\mcitedefaultendpunct}{\mcitedefaultseppunct}\relax
\EndOfBibitem
\bibitem[Sun \latin{et~al.}(2015)Sun, Ruzsinszky, and Perdew]{Sun:prl15} Sun,~J.; Ruzsinszky,~A.; Perdew,~J.~P. Strongly Constrained and Appropriately Normed Semilocal Density Functional. \emph{Phys. Rev. Lett.} \textbf{2015}, \emph{115}, 036402\relax
\mciteBstWouldAddEndPuncttrue
\mciteSetBstMidEndSepPunct{\mcitedefaultmidpunct}
{\mcitedefaultendpunct}{\mcitedefaultseppunct}\relax
\EndOfBibitem
\bibitem[Bokdam \latin{et~al.}(2017)Bokdam, Lahnsteiner, Ramberger, Sch\"afer, and Kresse]{Bokdam:prl17} Bokdam,~M.; Lahnsteiner,~J.; Ramberger,~B.; Sch\"afer,~T.; Kresse,~G. Assessing Density Functionals Using Many Body Theory for Hybrid Perovskites. \emph{Phys. Rev. Lett.} \textbf{2017}, \emph{119}, 145501\relax
\mciteBstWouldAddEndPuncttrue
\mciteSetBstMidEndSepPunct{\mcitedefaultmidpunct}
{\mcitedefaultendpunct}{\mcitedefaultseppunct}\relax
\EndOfBibitem
\bibitem[Lahnsteiner \latin{et~al.}(2018)Lahnsteiner, Kresse, Heinen, and Bokdam]{Lahnsteiner:prm18} Lahnsteiner,~J.; Kresse,~G.; Heinen,~J.; Bokdam,~M. Finite-temperature structure of the ${\mathrm{MAPbI}}_{3}$ perovskite: Comparing density functional approximations and force fields to experiment. \emph{Phys. Rev. Mat.} \textbf{2018}, \emph{2}, 073604\relax
\mciteBstWouldAddEndPuncttrue
\mciteSetBstMidEndSepPunct{\mcitedefaultmidpunct}
{\mcitedefaultendpunct}{\mcitedefaultseppunct}\relax
\EndOfBibitem
\bibitem[Sun \latin{et~al.}(2010)Sun, Shen, and Allen]{Sun:prb10} Sun,~T.; Shen,~X.; Allen,~B.,~Philip Phonon quasiparticles and anharmonic perturbation theory tested by molecular dynamics on a model system. \emph{Phys. Rev. B} \textbf{2010}, \emph{82}, 224304--1--9\relax
\mciteBstWouldAddEndPuncttrue
\mciteSetBstMidEndSepPunct{\mcitedefaultmidpunct}
{\mcitedefaultendpunct}{\mcitedefaultseppunct}\relax
\EndOfBibitem
\bibitem[Zhang \latin{et~al.}(2017)Zhang, Allen, Sun, and Wentzcovitch]{Zhang:prb17} Zhang,~D.-B.; Allen,~P.~B.; Sun,~T.; Wentzcovitch,~R.~M. Thermal conductivity from phonon quasiparticles with subminimal mean free path in the ${\mathrm{MgSiO}}_{3}$ perovskite. \emph{Phys. Rev. B} \textbf{2017}, \emph{96}, 100302(R)\relax
\mciteBstWouldAddEndPuncttrue
\mciteSetBstMidEndSepPunct{\mcitedefaultmidpunct}
{\mcitedefaultendpunct}{\mcitedefaultseppunct}\relax
\EndOfBibitem
\bibitem[Sun \latin{et~al.}(2014)Sun, Zhang, and Wentzcovitch]{Sun:prb14} Sun,~T.; Zhang,~D.-B.; Wentzcovitch,~R. Dynamic stabilization of cubic CaSiO3 perovskite at high temperatures and pressures from ab-initio molecular dynamics. \emph{Phys. Rev. B} \textbf{2014}, \emph{89}, 094109--1--9\relax
\mciteBstWouldAddEndPuncttrue
\mciteSetBstMidEndSepPunct{\mcitedefaultmidpunct}
{\mcitedefaultendpunct}{\mcitedefaultseppunct}\relax
\EndOfBibitem
\bibitem[Lahnsteiner and Bokdam(2021)Lahnsteiner, and Bokdam]{DSLEAP22} Lahnsteiner,~J.; Bokdam,~M. Dynamic Solids Large Ensemble Analysis Package (\textsc{dsleap}). \url{https://github.com/Jonathan271828/DSLEAP}, 2021; Accessed: March 1st 2023\relax
\mciteBstWouldAddEndPuncttrue
\mciteSetBstMidEndSepPunct{\mcitedefaultmidpunct}
{\mcitedefaultendpunct}{\mcitedefaultseppunct}\relax
\EndOfBibitem
\bibitem[Bussi and Parrinello(2007)Bussi, and Parrinello]{Bussi:pre07} Bussi,~G.; Parrinello,~M. Accurate sampling using Langevin dynamics. \emph{Phys. Rev. E} \textbf{2007}, \emph{75}, 056707\relax
\mciteBstWouldAddEndPuncttrue
\mciteSetBstMidEndSepPunct{\mcitedefaultmidpunct}
{\mcitedefaultendpunct}{\mcitedefaultseppunct}\relax
\EndOfBibitem
\bibitem[Student(1908)]{student1908probable} Student, The probable error of a mean. \emph{Biometrika} \textbf{1908}, \emph{6}, 1--25\relax
\mciteBstWouldAddEndPuncttrue
\mciteSetBstMidEndSepPunct{\mcitedefaultmidpunct}
{\mcitedefaultendpunct}{\mcitedefaultseppunct}\relax
\EndOfBibitem
\bibitem[Simoncelli \latin{et~al.}(2022)Simoncelli, Marzari, and Mauri]{Simoncelli:prx22} Simoncelli,~M.; Marzari,~N.; Mauri,~F. Wigner Formulation of Thermal Transport in Solids. \emph{Phys. Rev. X} \textbf{2022}, \emph{12}, 041011\relax
\mciteBstWouldAddEndPuncttrue
\mciteSetBstMidEndSepPunct{\mcitedefaultmidpunct}
{\mcitedefaultendpunct}{\mcitedefaultseppunct}\relax
\EndOfBibitem
\bibitem[Kanak \latin{et~al.}(2020)Kanak, Lishchuk, Kuryliuk, Kuzmich, Lacroix, Khalavka, and Isaiev]{Kanak:aip20} Kanak,~A.; Lishchuk,~P.; Kuryliuk,~V.; Kuzmich,~A.; Lacroix,~D.; Khalavka,~Y.; Isaiev,~M. Thermal conductivity of CsPbBr3 halide perovskite: Photoacoustic measurements and molecular dynamics analysis. \emph{AIP Conference Proceedings} \textbf{2020}, \emph{2305}, 020006\relax
\mciteBstWouldAddEndPuncttrue
\mciteSetBstMidEndSepPunct{\mcitedefaultmidpunct}
{\mcitedefaultendpunct}{\mcitedefaultseppunct}\relax
\EndOfBibitem
\bibitem[S\"a\"askilahti \latin{et~al.}(2014)S\"a\"askilahti, Oksanen, Tulkki, and Volz]{Saaskilahti:prb14} S\"a\"askilahti,~K.; Oksanen,~J.; Tulkki,~J.; Volz,~S. Role of anharmonic phonon scattering in the spectrally decomposed thermal conductance at planar interfaces. \emph{Phys. Rev. B} \textbf{2014}, \emph{90}, 134312\relax
\mciteBstWouldAddEndPuncttrue
\mciteSetBstMidEndSepPunct{\mcitedefaultmidpunct}
{\mcitedefaultendpunct}{\mcitedefaultseppunct}\relax
\EndOfBibitem
\bibitem[Lv and Henry(2016)Lv, and Henry]{Lv:njp16} Lv,~W.; Henry,~A. Direct calculation of modal contributions to thermal conductivity via Green–Kubo modal analysis. \emph{New J. Phys.} \textbf{2016}, \emph{18}, 013028\relax
\mciteBstWouldAddEndPuncttrue
\mciteSetBstMidEndSepPunct{\mcitedefaultmidpunct}
{\mcitedefaultendpunct}{\mcitedefaultseppunct}\relax
\EndOfBibitem
\bibitem[Caddeo \latin{et~al.}(2016)Caddeo, Melis, Saba, Filippetti, Colombo, and Mattoni]{Caddeo:pccp16} Caddeo,~C.; Melis,~C.; Saba,~M.~I.; Filippetti,~A.; Colombo,~L.; Mattoni,~A. Tuning the thermal conductivity of methylammonium lead halide by the molecular substructure. \emph{Phys. Chem. Chem. Phys.} \textbf{2016}, \emph{18}, 24318--24324\relax
\mciteBstWouldAddEndPuncttrue
\mciteSetBstMidEndSepPunct{\mcitedefaultmidpunct}
{\mcitedefaultendpunct}{\mcitedefaultseppunct}\relax
\EndOfBibitem
\bibitem[Bokdam \latin{et~al.}(2021)Bokdam, Lahnsteiner, and Sarma]{Bokdam:jpcc21} Bokdam,~M.; Lahnsteiner,~J.; Sarma,~D.~D. Exploring Librational Pathways with on-the-Fly Machine-Learning Force Fields: Methylammonium Molecules in MAPbX3 (X = I, Br, Cl) Perovskites. \emph{J. Phys. Chem. C} \textbf{2021}, \emph{125}, 21077--21086\relax
\mciteBstWouldAddEndPuncttrue
\mciteSetBstMidEndSepPunct{\mcitedefaultmidpunct}
{\mcitedefaultendpunct}{\mcitedefaultseppunct}\relax
\EndOfBibitem
\bibitem[Bokdam \latin{et~al.}(2023)Bokdam, Lahnsteiner, and Rang]{Bokdam:4TUDATA23} Bokdam,~M.; Lahnsteiner,~J.; Rang,~M. CsPbBr3 database for force field training underlying the publication: Tuning Einstein Oscillator Frequencies of Cation Rattlers: A Molecular Dynamics Study of the Lattice Thermal Conductivity of CsPbBr3. 2023; \url{https://doi.org/10.4121/02c647b6-c6cb-4e45-b7d5-287853152c47.v1}\relax
\mciteBstWouldAddEndPuncttrue
\mciteSetBstMidEndSepPunct{\mcitedefaultmidpunct}
{\mcitedefaultendpunct}{\mcitedefaultseppunct}\relax
\EndOfBibitem
\bibitem[Kuipers and Bokdam(2023)Kuipers, and Bokdam]{FPdataViewer23} Kuipers,~T.; Bokdam,~M. FPdataViewer: factsheets for ensembles of electronic structure data. \url{https://github.com/dynamicsolids/FPdataViewer}, 2023\relax
\mciteBstWouldAddEndPuncttrue
\mciteSetBstMidEndSepPunct{\mcitedefaultmidpunct}
{\mcitedefaultendpunct}{\mcitedefaultseppunct}\relax
\EndOfBibitem
\end{mcitethebibliography}
\end{document}


\maketitle
\tableofcontents

\section{The Machine Learning Force Fields}
\subsection{Training}
To measure the thermal conductivity of the CsPbBr$_{3}$ pervoskite, molecular dynamics  simulations at 500~K were employed. The molecular dynamics calculations were performed with machine-learning force fields (MLFF) as implemented in VASP~\cite{Jinnouchi:prl19,Jinnouchi:prb19,Jinnouchi:jcp20}.
We are interested in the thermal conductivity for CsPbBr$_{3}$ with 3 different masses. Eventhough the DFT potential does not depend on the mass, the dynamics explored during the on-the-fly training is affected. Therefore, we train three force fields, one for every mass. The training procedure is the same for every Cs mass. The SCAN exchange correlation functional\cite{Sun:prl15} was shown in the past
to give reliable results for this type of material~\cite{Jinnouchi:prl19,Lahnsteiner:prm18,Bokdam:prl17}.
Therefore we use the SCAN functional~\cite{Sun:prl15} with an energy cutoff of 287~eV. We start the training procedure at 50~K with an equilibrated orthorhombic crystal structure in a $2\times2\times2$ formula-unit supercell. This structure is heated during an on-the fly training run starting at 50~K to 500~K in the NPT ensemble.
A time step of $5$~fs was used and 40,000 steps were simulated in total, resulting in a temperature gradient of $2.25$~K/ps. The external stress was set to 0.001~kbar. The cutoff radii for the radial and angular descriptors were set to 5~$\AA$. The number of spherical Bessel functions for the radial part and the angular part of the descriptor were set to 8. The maximum angular momentum number for the expansion of the angular part was adjusted to $l_{max}=4$. After the training during heating was done, another training round is done at 500~K. For the high temperature training we took the last structure of the heating run and propagated this structure for 100~ps. During training we were using Bayes linear regression to
determine the regression coefficients. Before the production runs we recomputed the regression coefficients of our force fields with singular value decomposition (SVD). This procedure was reported to give lower root mean square errors (RMSE) in the energy forces and stress
tensor~\cite{Verdi:npjcm21}. The number of used reference configurations for the different atoms can be found in table~\ref{refconftable}

\begin{table}[!h]
        \centering
        
   \caption{Number of reference configurations in the various force fields }
        \begin{tabular}{ | c | c | c | c |}
                \hline
      Cs mass & Pb & Br & Cs \\
                \hline
        13.29 & 204 & 940 & 173 \\
        132.9 & 165 & 960 & 135 \\
        1329  & 163 & 885 & 133 \\
                \hline 
        \end{tabular} 
   \label{refconftable}
   
\end{table}

\newpage

\subsection{Error analysis}
\begin{figure}[!h]
	\centering
	\includegraphics[width=.6\linewidth]{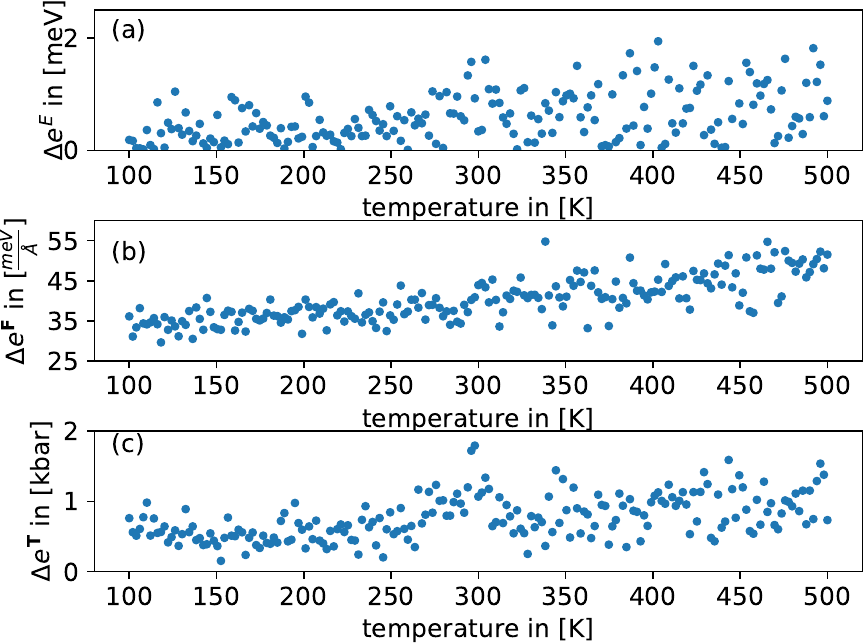}
	\caption{Error analysis of MLFF trained with $m_{Cs}=13.29054$.
		 \textbf{a:} Error in total energy per atom. \textbf{b:} RMSE in of the force.
		 \textbf{c:} RMSE of the stress tensor. The x-axis is the
		 temperature at which the test structure was taken.}
	\label{fig:error_light}
\end{figure}

\begin{figure}[!h]
	\centering
	\includegraphics[width=.6\linewidth]{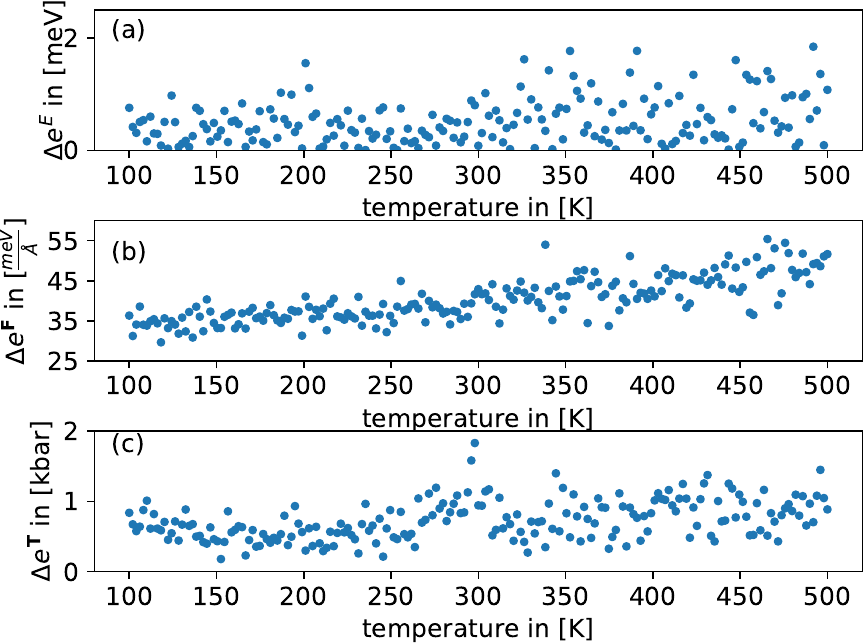}
	\caption{Error analysis of MLFF trained with $m_{Cs}=132.9054$.
		 \textbf{a:} Error in total energy per atom. \textbf{b:} RMSE in of the force.
		 \textbf{c:} RMSE of the stress tensor. The x-axis is the
		 temperature at which the test structure was taken.}
	\label{fig:error_normal}
\end{figure}

\begin{figure}[!h]
	\centering
	\includegraphics[width=.6\linewidth]{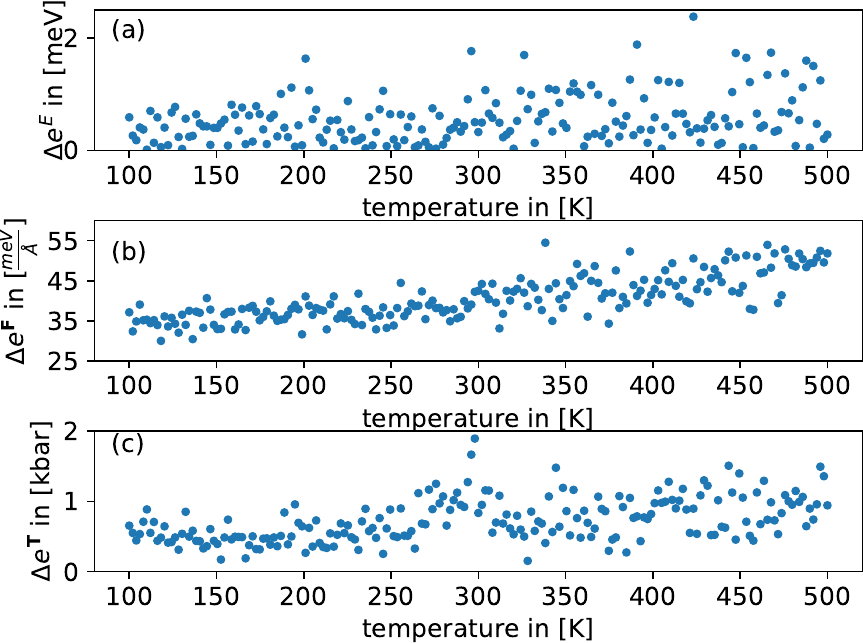}
	\caption{Error analysis of MLFF trained with $m_{Cs}=1329.054$.
		 \textbf{a:} Error in total energy per atom. \textbf{b:} RMSE in of the force.
		 \textbf{c:} RMSE of the stress tensor. The x-axis is the
		 temperature at which the test structure was taken.}
	\label{fig:error_heavy}
\end{figure}

To quantify the error in the energy, the forces and the stress tensor of the MLFF a set of independent test structures was created. To create
a set of test structures the CsPbr$_{3}$ force field of Refs.~\cite{Jinnouchi:prl19,Lahnsteiner:prb22} was used.
With this force field test structures were created during a heating run, starting from 50~K and going up to 500~K. The temperature gradient was adjusted as in the training procedure. From this heating run 200 equidistant structure files were extracted, giving a POSCAR file
after every 1~ps or every 2.25~K. For this 200 structures the energy, forces and stress tensor were computed with the SCAN functional and the force-fields. The average energy error,

\begin{equation}
        \Delta e^{E} =\frac{|E_{DFT}-E_{ML}|}{N}
        \label{EnergyErrorY}
\end{equation}
\noindent
was computed according to an absolute energy difference divided by the number of atoms. The error in the forces is computed as a RMSE,
\begin{equation}
	\Delta e^{\mathbf{F}} = \sqrt{\frac{1}{3N}\sum_{i=1}^{N}
	          \sum_{\alpha=1}^{x,y,z} \left ( F_{i,\alpha}^{DFT} - F_{i,\alpha}^{ML} \right)^{2}},
        \label{ForceErrorY}
\end{equation}
\noindent
where the index $\alpha$ is running over the Cartesian components of the force vector and the index $i$ runs over all the atoms in the supercell.
The force vector is plotted against the average magnitude of the DFT force of the considered structure. The RMSE in the stress tensor $\mathbf{T}$ is computed over the 6 independent components of the tensor, 
\begin{equation}
	\Delta e^{\mathbf{T}} = \sqrt{\frac{1}{6}\sum_{\alpha=1}^{6} \left ( T_{\alpha}^{DFT} - T_{\alpha}^{ML} \right)^{2}}.
        \label{StressErrorY}
\end{equation}
\noindent
The errors for the different Cs mass force-fields are summarized in Fig.~\ref{fig:error_light} for the light
mass, in Fig.~\ref{fig:error_normal} for the normal mass and the error of the heavy mass is shown in Fig.~\ref{fig:error_heavy}.\newpage

\newpage

\newpage
\section{q-VACF}
\begin{figure}[!h]
	\centering
	\includegraphics[width=.7\linewidth]{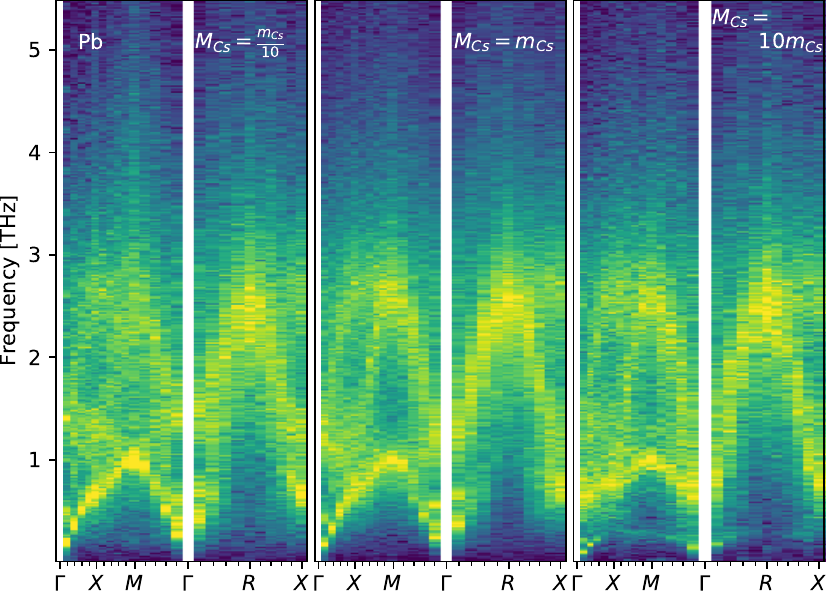}
	\caption{Pb contribution to the power-spectrum of the $\mathbf{q}$-VACF for varying Cs mass. The mass increases from left to right.}
	\label{fig:qvacfPb}
\end{figure}

\begin{figure}[!h]
	\centering
	\includegraphics[width=.7\linewidth]{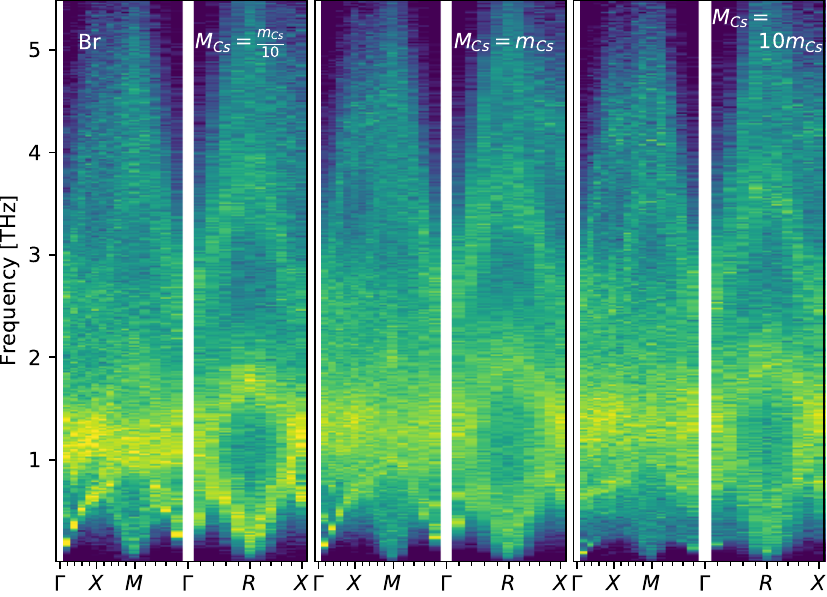}
	\caption{Br contribution to the power-spectrum of the $\mathbf{q}$-VACF for varying Cs mass. The mass increases from left to right}
	\label{fig:qvacfBr}
\end{figure}

\begin{figure}[!h]
	\centering
	\includegraphics[width=.7\linewidth]{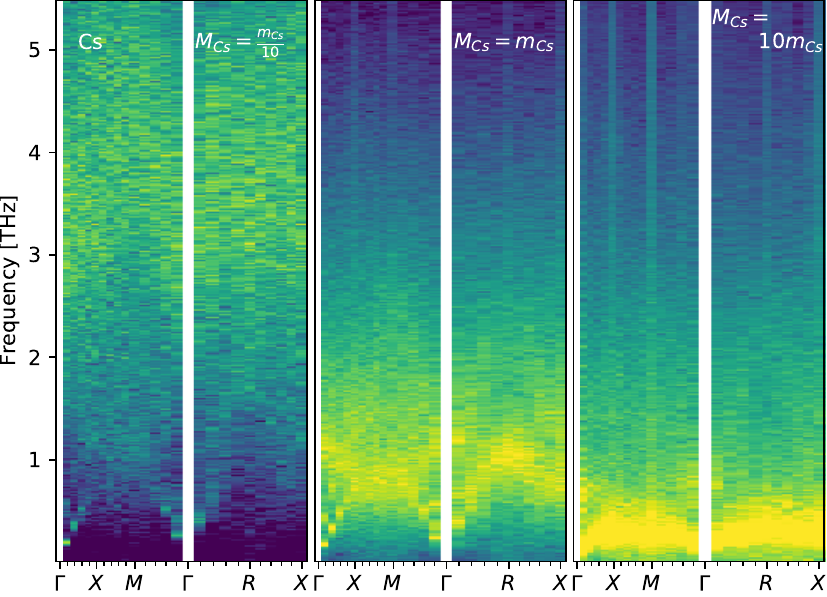}
	\caption{Cs contribution to the power-spectrum of the $\mathbf{q}$-VACF for varying Cs mass. The mass increases from left to right}
	\label{fig:qvacfCs}
\end{figure}

The dynamic properties of the crystal are accessed with the 
$\mathbf{q}$-resolved velocity-autocorrelation-function ($\mathbf{q}$-VACF). The $\mathbf{q}$-point is one of the commensurate reciprocal grid points. In this manner we are able to qualitatively determine how a change of the Cs mass affect the phonon band structure. We compute the $\mathbf{q}$-VACF by Fourier transforming the 
mass weighed velocity field to $\mathbf{q}$-space
\begin{equation}
        \tilde{\mathbf{v}}_{s}(\mathbf{q},t)=\sum_{\mathbf{n}}\sqrt{m}_{s}\mathbf{v}_{s}
              (\mathbf{n},t)e^{i\mathbf{q}\mathbf{r}_{s}(\mathbf{n},t)},
        \label{qVelocity}
\end{equation}
\noindent
Then Eq.~(\ref{qVelocity}) is self-correlated and the temporal Fourier transform is computed
\begin{equation}                                                    
        \tilde{g}_{s}(\mathbf{q},\omega)=\int \tilde{\mathbf{v}}_{s}(\mathbf{q},t')
             \tilde{\mathbf{v}}_{s}(\mathbf{-q},t'') e^{-i\omega t} dt,
        \label{Q-VACFdef}
\end{equation}                                                                    
\noindent
with $t=t'-t''$ and d$t$ the corresponding differential. The power spectrum of the so obtained function is a $\mathbf{q}$-resolved form of the phonon DOS~\cite{Lahnsteiner:prb22}. How the power-spectrum and the ensemble average of Eq.~(\ref{Q-VACFdef}) are computed is described in detail in Ref~\cite{Lahnsteiner:prb22}. The obtained results are shown in Figs.~\ref{fig:qvacfPb} to~\ref{fig:qvacfCs}.\newpage

\newpage
\section{Simulation setup of the active $\kappa$ measurement: Langevin friction and system size dependence}

\begin{figure}[!t]
	\centering
	\includegraphics[width=.72\linewidth]{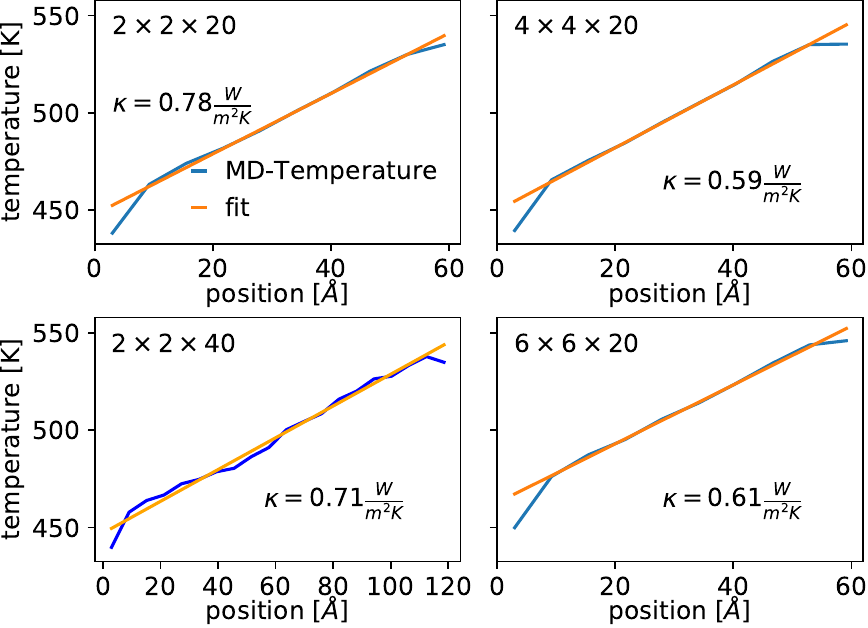}
	\caption{Convergence of the heat gradient for various system sizes in the active measurement approach using a too large Langevin friction (10~ps$^{-1}$). The system sizes and the calculated $\kappa$ are shown in the plots. }	
	\label{fig:SizeConvergence}
\end{figure}

To compute the thermal conductivity we used the method of active measurement with a constant temperature
gradient\cite{Zhou:prb19,Li:jcp19,Dunn:jap16}. A finite size analysis was carried out by simulating 1~ns long single trajectories in $2\times2\times20$, $2\times2\times40$, $4\times4\times20$ and $6\times6\times20$ cubic
supercells. The fixed temperature difference was set to 100~K resulting in $450$~K and $550$~K for the cold $\Gamma_{C}$ and hot $\Gamma_{H}$ region in the box respectively. These temperatures are all safely above the tetragonal-cubic phase transition temperature of ~380~K \cite{Jinnouchi:prl19} to ensure that the structure in the entire supercell remains in the cubic phase. A Langevin thermostat was placed only on the atoms in the $\Gamma_{H}$ and $\Gamma_{C}$ regions. We view these regions as leads that a fixed width of two lattice constants ($2a$), and enclose a scattering region in between them. \textbf{Oppositely to the simulations presented in the paper, the test simulations shown in Figure ~\ref{fig:SizeConvergence} have been performed with a 10$\times$ higher Langevin friction, ie. 10~ps$^{-1}$.}  The equilibrium starting structures were obtained from MD runs of 50~ps in an NVT ensemble. The resulting temperature profiles after averaging the temperature per layer over time are shown in Figure~\ref{fig:SizeConvergence}. Artefacts appear in the temperature profile near the thermostated leads. Whereas the jump up in temperature at the cold lead (position left) could still be interpreted as an interface thermal resistance, the jump down at the hot lead (position right) makes no physical sense. In an attempt to resolve this artefact we stay closer to the linear response regime, in which the leads cause a small perturbation out of equilibrium. Therefore, the temperature difference between the leads was lowered to $500\pm 25$~K, \textbf{followed by lowering the Langevin friction coefficient to 1~ps$^{-1}$. This removes the unphysical temperature jumps between the leads and the scattering region.} Knowing this, we attribute the artefact to the MLFF (as described in section SI 1.1) being inferred for local atomic environments that lie too far outside the phase space on which it was trained. It was trained at 500~K with a friction coefficient of 10~ps$^{-1}$, however the hot lead was set to 550~K, and the same friction was applied. By lowering both $\Delta{}T$ and the Langevin friction, the MLFF remains able to assign a correct local energy to the hot part of the simulation box.

\begin{figure}[!t]
	\centering
	\includegraphics[width=.49\linewidth]{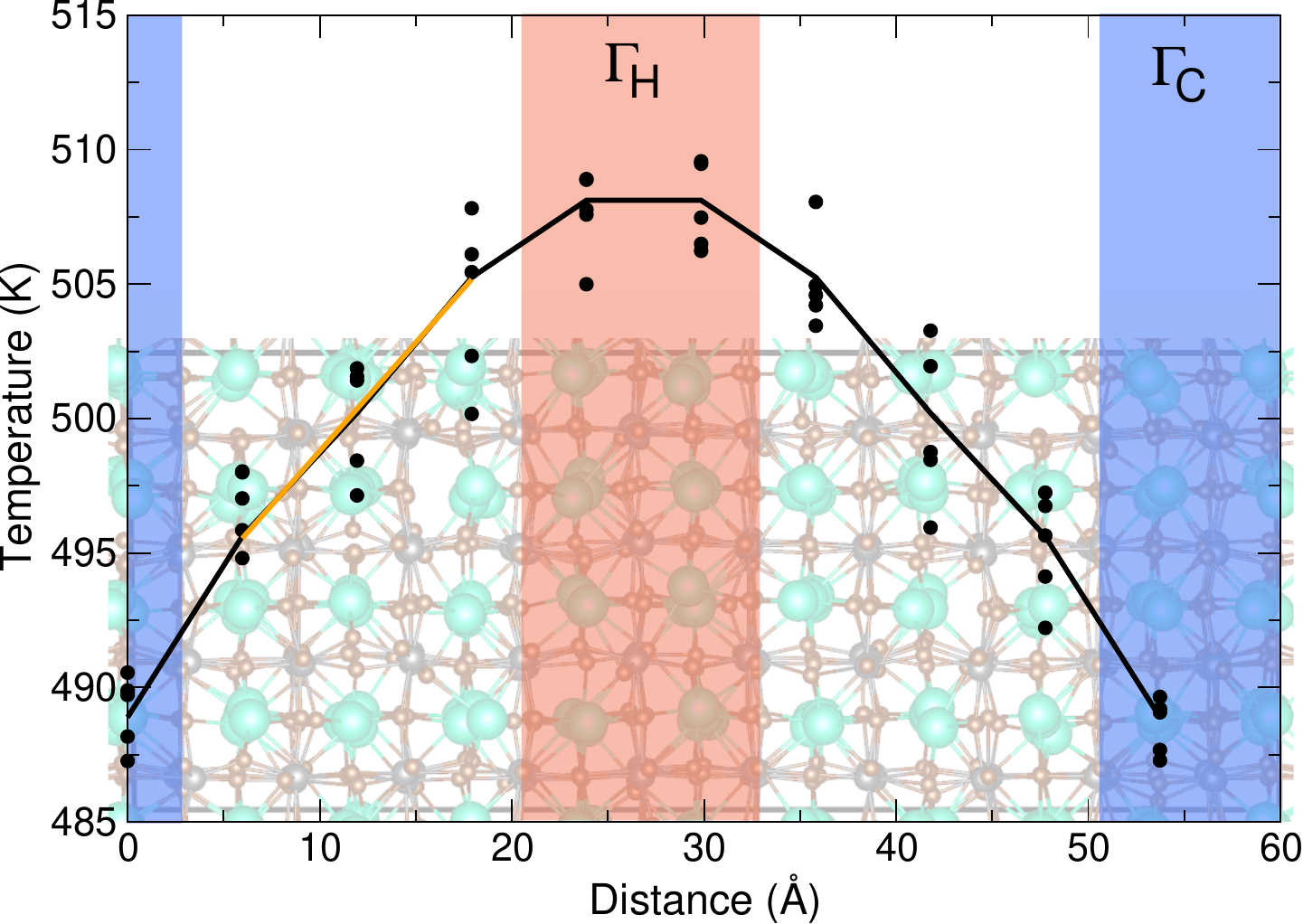}
 \includegraphics[width=.49\linewidth]{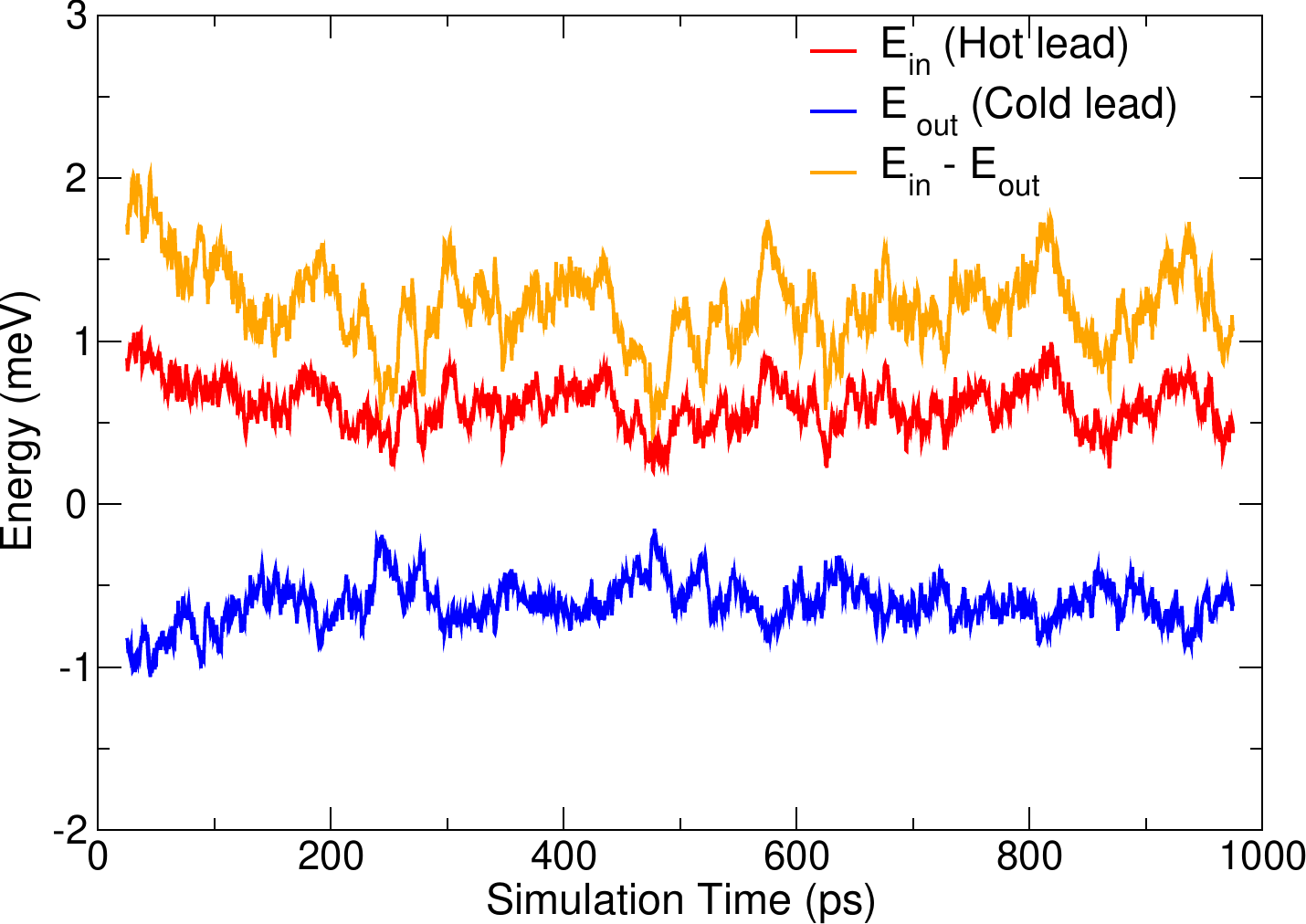}
	\caption{CsPbBr$_3$ (normal mass) NEMD simulations of a short $4\times4\times10$ supercell under periodic boundary conditions. \textit{Left:} Temperature profile, thermostats keep the (red) $\Gamma_{\rm H}$-layers at high and the (blue) $\Gamma_{\rm C}$-layers at low constant temperature. A stable temperature gradient develops in the time average of the layer temperatures, the orange line is the fit determining $\langle\nabla{}T\rangle$.  \textit{Right:} Energy balance during the NEMD trajectory visualised by a moving average with window size of 50~ps. In both the left and right figure the lines represent data averaged over five independent trajectories.}	
	\label{fig:tempgrad4410}
\end{figure}

\noindent From this point on, and in all simulations presented in the main text, a Langevin friction coefficient of 1~ps$^{-1}$ was used. We have observed that the variance in the $\kappa$ values from independent simulations in the same dimension supercell is more important, than the effect of a changing its size (Fig.~\ref{fig:SizeConvergence}). Since the expected mean free path of almost all phonon modes at 500~K is below 10$\AA$ (see discussion in main text), a simulation box with a scattering region length of $L=3a=\sim18\AA$ (region in which no thermostat is applied) can start to approach the bulk thermal conductivity. For this small supercell the temperature profile that develops on average over 1~ns of MD simulation is shown in Figure~\ref{fig:tempgrad4410}(left). In the energy balance of Fig.~\ref{fig:tempgrad4410}(right) we see that after $\sim200$~ps the NEMD simulation reaches a steady state. On average, the hot lead pumps in as much kinetic energy as the cold lead extracts from the system. In Table~\ref{tbl:4410} we have compared the $\kappa$ values obtained from 5 independent simulations of a $4\times4\times10$ supercell ($L=3a$), to the values presented in the main text for a $4\times4\times20$ supercell ($L=8a$). What we see is that the mean value is practically the same, but the confidence interval is higher in the smaller supercell.\vspace{3mm}

\begin{table}[t!]
	\caption{Lattice thermal conductivity ($\kappa$) in $\frac{\rm W}{\rm m K}$ for CsPbBr$_3$ at 500~K with normal Cs masses. Results form a $4\times4\times10$ and a $4\times4\times20$ supercell, with length of the scattering region of $3a$ and $8a$, respectively.}
	\begin{tabular}{|m{0.35\columnwidth}|m{0.15\columnwidth}|m{0.15\columnwidth}|m{0.15\columnwidth}|}\hline
		   &  L=3a   & L=8a  \\
	\hline
	Run 1 &  0.47 & 0.51 \\
	Run 2 &  0.41 & 0.50 \\
	Run 3 &  0.62 & 0.63 \\
	Run 4 &  0.55 & 0.55 \\
	Run 5 &   & 0.54 \\
	\hline
	Mean  &  0.51 & 0.55 \\
 	Confidence interval &$\pm$0.096 & $\pm$0.042 \\
    \hline
	\end{tabular}
	\label{tbl:4410}
\end{table}

\begin{figure}[h!]
	\centering
  	\includegraphics[width=.49\linewidth]{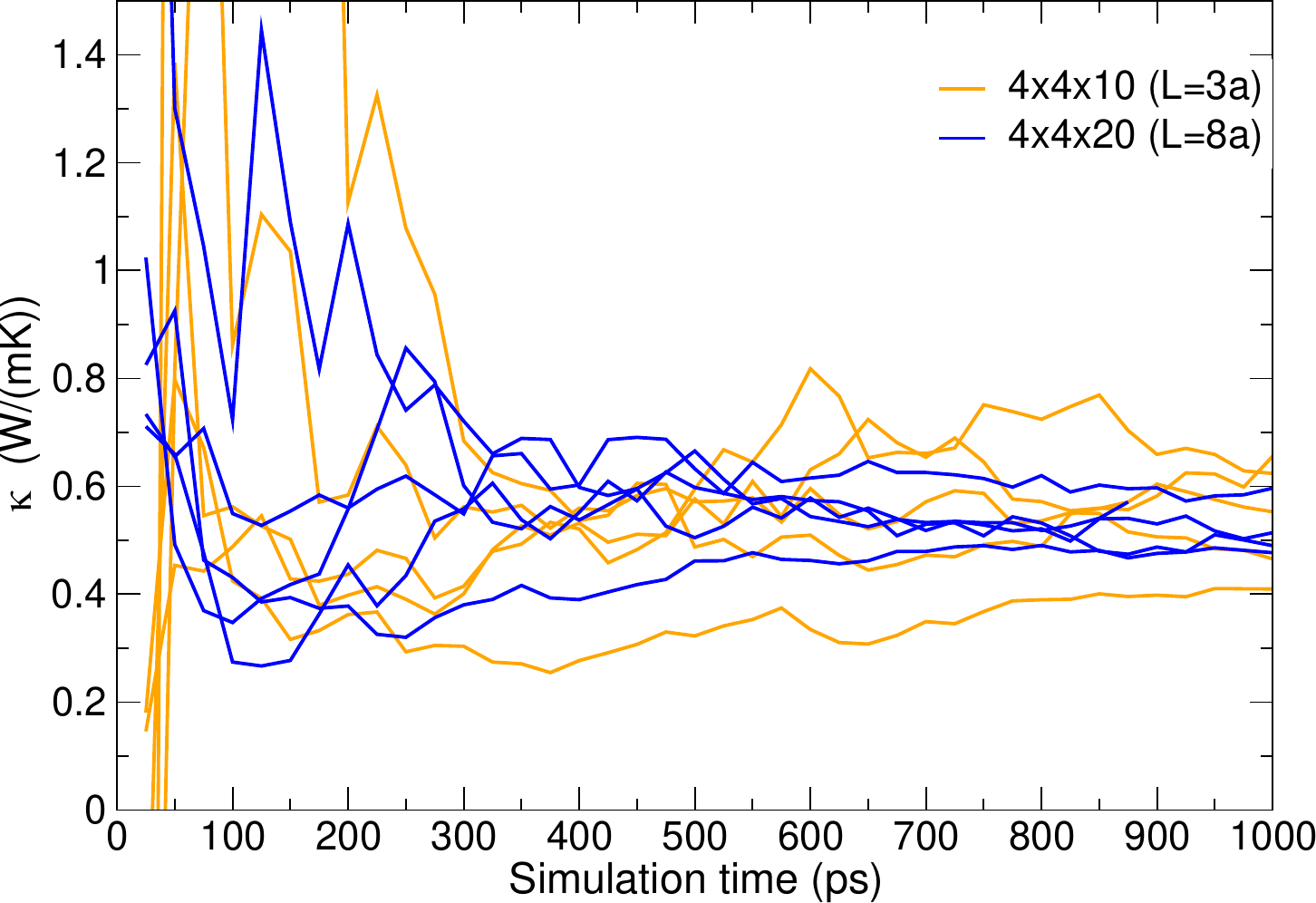}
	\includegraphics[width=.49\linewidth]{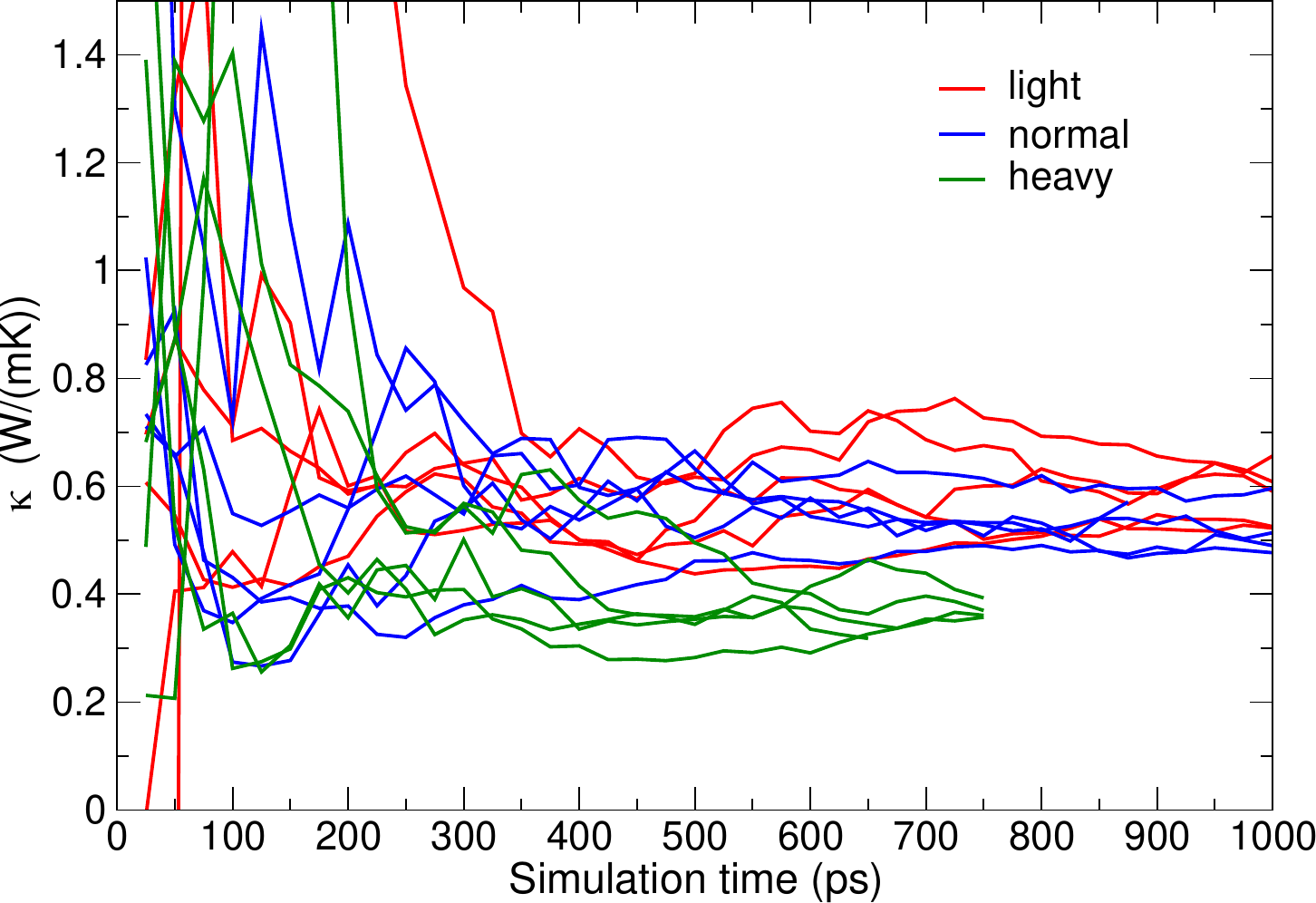}

	\caption{Convergence of $\kappa$ with simulation time in active measurement approach for 5 independent trajectories per system configuration. \textit{Left:} A comparison of $\kappa$ in the $4\times4\times20$ ($L=8a$) and in a smaller $4\times4\times10$ ($L=3a$) system.  \textit{Right:} $\kappa$ vs. simulation time for  different Cs masses: light, normal and heavy.  }
	\label{fig:KappaConvergence}
\end{figure}

The convergence of the thermal conductivity with simulation time is demonstrated in Figure~\ref{fig:KappaConvergence} for the individual trajectories. The ensemble averages in Fourier's law, $
	\kappa=\frac{-\langle\Delta Q\rangle}{2\langle\nabla{}T\rangle} ,$ are evaluated for an increasing part of the in total near 1~ns trajectories. In the left figure we convergence of $\kappa$ in the $L=3a$ (orange lines) is compared to $L=8a$ (blue lines). The values in Table~\ref{tbl:4410} are the $\kappa$ values at the ends of the curves. In the right figure the mass of the Cs cations was varied. Here, all trajectories are for $4\times4\times20$ supercells. What we see is that $\kappa$ values obtained from simulations starting from snapshots of the NPT ensemble converge within a uncertainty interval. This interval is narrow enough such that we can conclude that on average $\kappa_{\rm light}>\kappa_{\rm normal}>\kappa_{\rm heavy}$.

\newpage

\section{Lorentzian fitting procedure PVACF spectra}
Figure 6 in the manuscript shows PVACF spectra where we projected the velocity field onto the eigenvector of the first acoustic mode ($\alpha=1$). Spectra   $P(\omega)$ corresponding to $\mathbf{q}$-points on the $\Gamma-$X line in reciprocal space are fitted with a Lorentzian function
\begin{equation}
 I_{\Gamma,\omega_0,I_0}(\omega)=I_0 \frac{\Gamma}{(\omega-\omega_0)^2+\Gamma^2}.
\end{equation}
We obtained \{$\Gamma,\omega_0,I_0$\}-values by minimizing the $L_1$-norm
\begin{equation}
 {\left|P(x)-I_{\Gamma,\omega_0,I_0}(x)\right|}_1=\int_0^{\omega_{\rm max}} |P(\omega)-I_{\Gamma,\omega_0,I_0}(\omega)| d\omega.
\end{equation}
In this way we approximate the spectrum by the best-fitting single broadened oscillator.

\begin{figure}[!h]
	\centering
	\includegraphics[width=.7\linewidth]{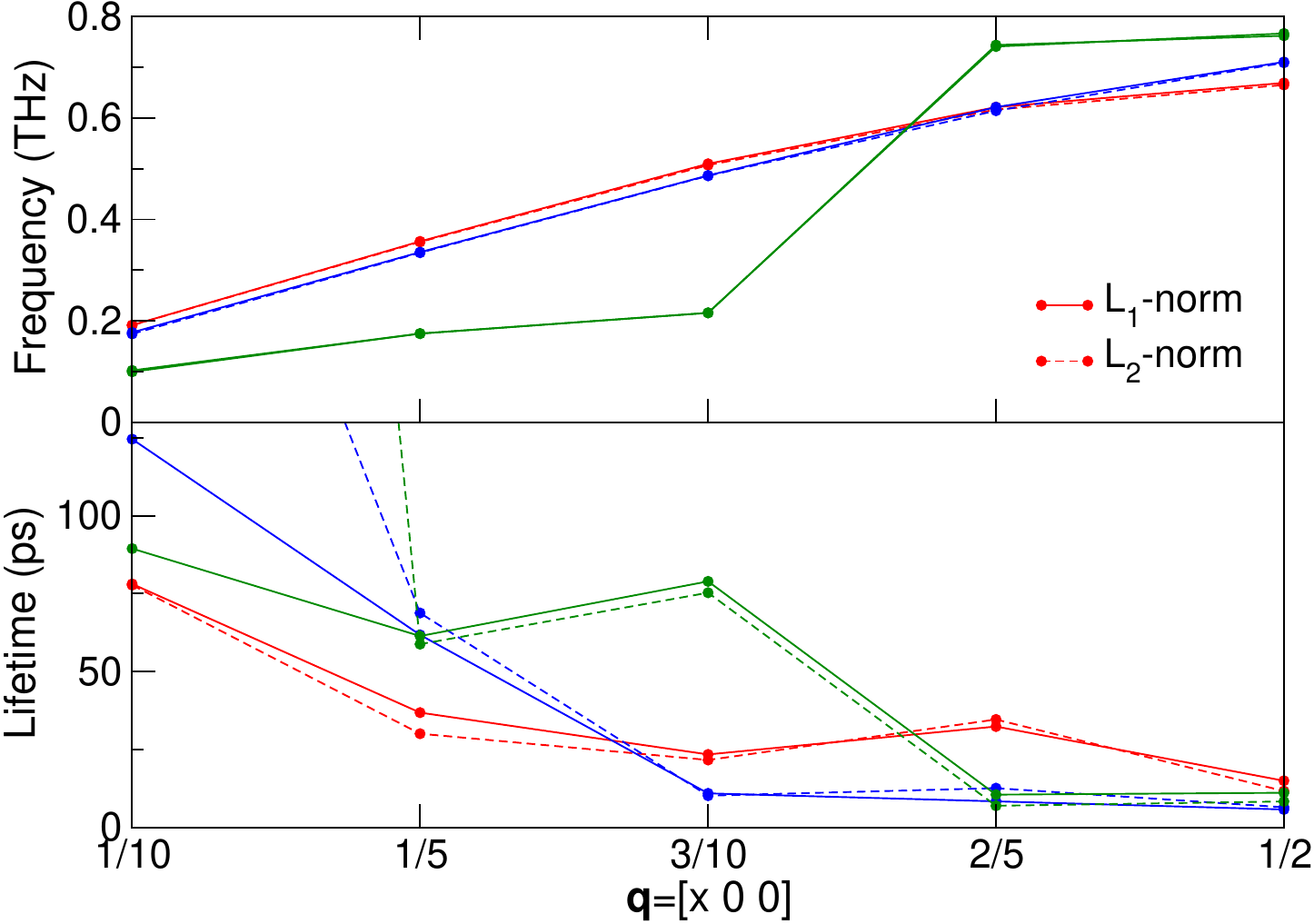}
	\caption{Parameters of fitted Lorentzians: mean frequency ($\omega_0$) and lifetime ($\frac{1}{2\Gamma}$) for the first acoustic mode ($\alpha=1$). The solid/dashed lines correspond to optimizing w.r.t. $L_1/L_2$-norm, respectively. }
	\label{fig:L1L2}
\end{figure}

We have tested whether $L_2$-norm affects the fitted \{$\Gamma,\omega_0$\}-values. Figure~\ref{fig:L1L2} shows that the $L_2$-norm results in the same frequency ($\omega_0$) and, the same lifetimes ($\frac{1}{2\Gamma}$). Only at $\mathbf{q}=[1/10\; 0\; 0]$ the coarseness of the frequency grid combined with the narrow peak width makes that $\Gamma$ depends on the applied norm. In case of the heavy and the normal mass system, the $L_2$-norm results in a more narrow Lorentzian and therefore a higher lifetime (heavy mass: 1000~ps, normal mass: 303~ps).

\begin{figure*}
	\centering
	\begin{tabular}{p{.49\linewidth}p{.49\linewidth}}
	 $\mathbf{q}=[\frac{1}{10} 0 0]$& $\mathbf{q}=[\frac{4}{10} 0 0]$\\
	 \includegraphics[width=\linewidth]{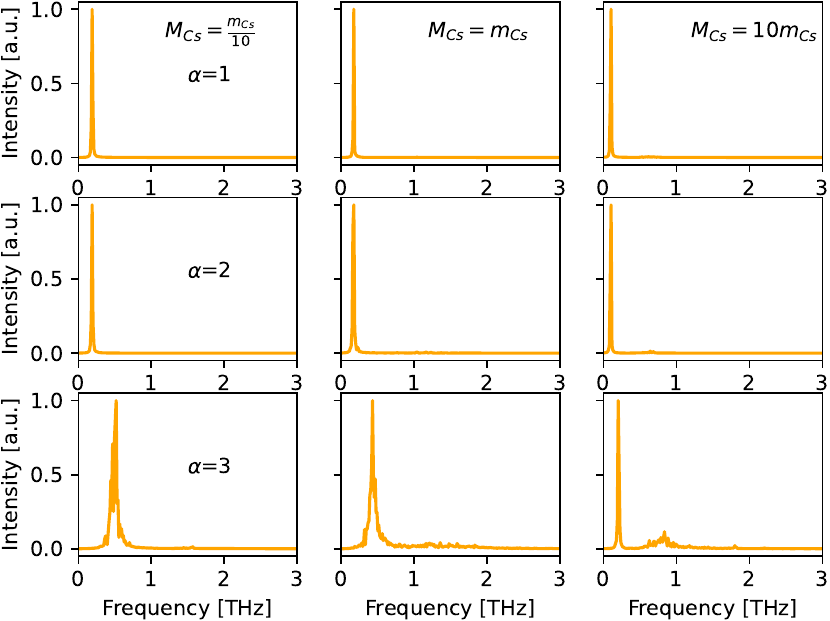}&
	 \includegraphics[width=\linewidth]{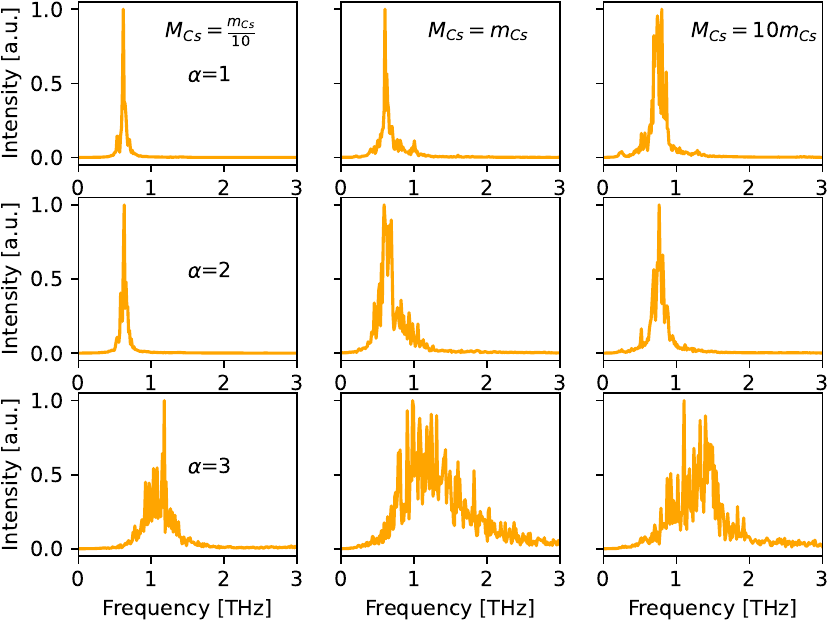} \\
	 	 $\mathbf{q}=[\frac{2}{10} 0 0]$& $\mathbf{q}=[\frac{5}{10} 0 0]$\\
	 \includegraphics[width=\linewidth]{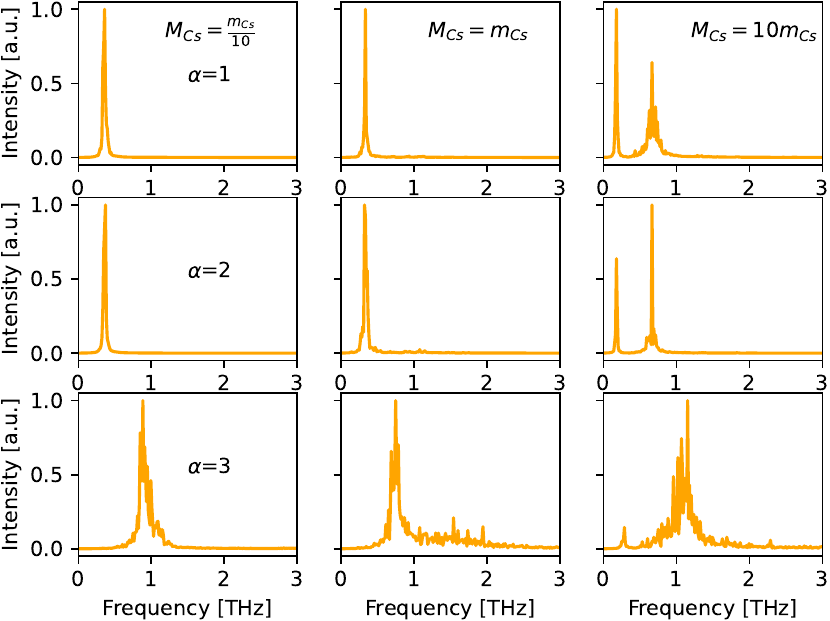}&
	 \includegraphics[width=\linewidth]{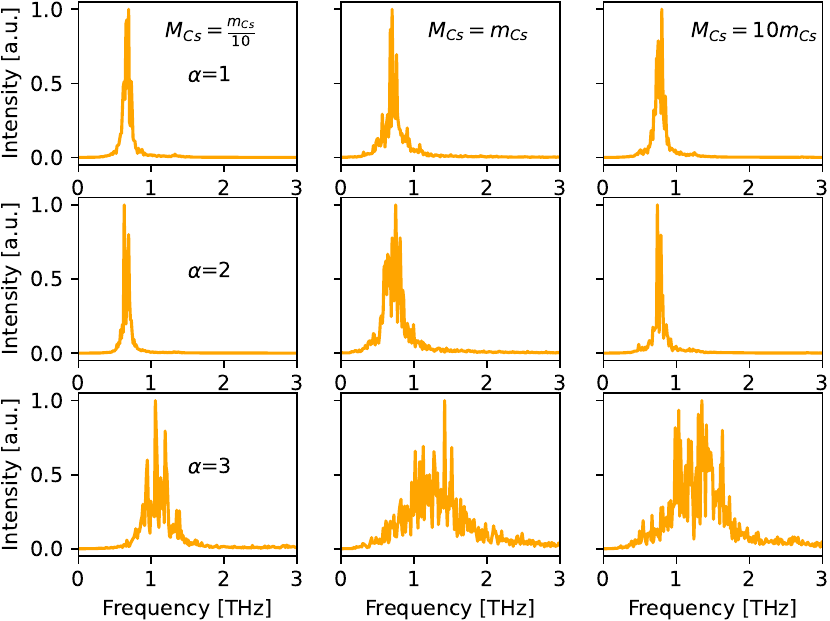} \\
	 	 $\mathbf{q}=[\frac{3}{10} 0 0]$& \\
	 \includegraphics[width=\linewidth]{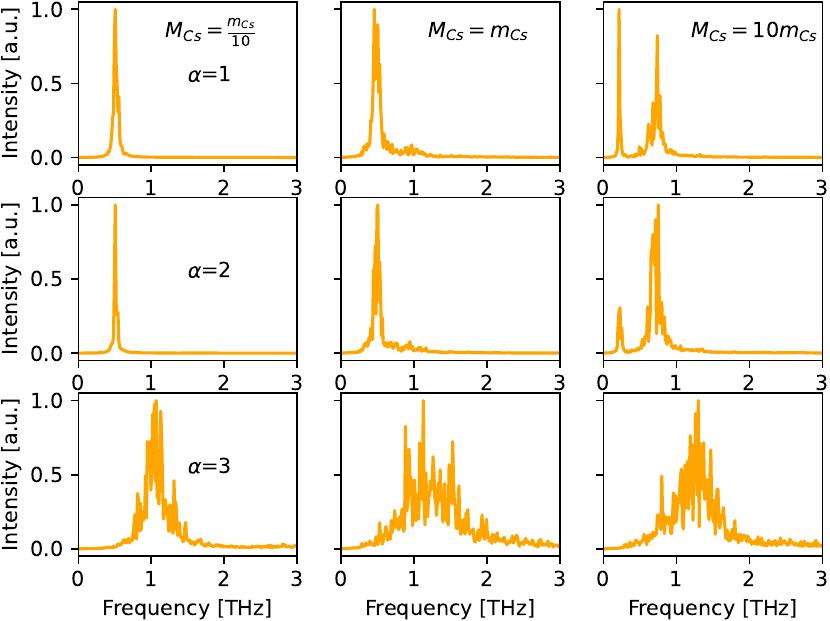}&
	  \\
	\end{tabular}

	\caption{PVACF spectra at $\mathbf{q}$-points on the $\Gamma$-X line for the low, normal and high Cs mass, and the three acoustic modes $\alpha=1,2,3$.}
	\label{fig:spectra}
\end{figure*}

For comparison with Figure 6 in the manuscript, we include in Figure~\ref{fig:spectra} the PVACF spectra projected onto the eigenvector of the first, second and third acoustic mode ($\alpha=1,2,3$). It shows that a similar plot as Figure~6(a) in the manuscript can be made for the $\alpha=2$ mode. The multi-peak spectra observed for the $\alpha=3$ mode become very broad.

\section{Calculation of Peierls' conductivity}
The extracted phonon quasi-particle properties for the first acoustic mode ($\alpha=1$) based on Lorentzian fitting as explained in the previous section are shown in Table~\ref{table:param}. The group velocity $\mathbf{v}_\alpha{}(\mathbf{q})=\nabla_\mathbf{q}\omega_\alpha(\mathbf{q})$ is computed by numerical differentiation. The central differences scheme is applied on the interior points on the line $\Gamma-{\rm X}$, and the Euler forward/backward differentiation is applied at the start/end points, respectively. The occupancy of the mode $n$ is determined by the height of the Lorentzian and is expressed as a relative number. The same holds for the group velocity, ie. prefactors are not taken into account. The absolute value of the resulting Peierls' conductivity (Eq.(1) in the main text) has therefore an arbitrary prefactor, but ratio's between the 15 numbers in Table~\ref{table:param} are correct. To indicate this, the $\kappa_{\rm P}$ values in Figure~7 of the main text have been expressed in percentages.

\begin{table}
 \caption{Extracted phonon quasi-particle properties for the first acoustic mode ($\alpha=1$) based on Lorentzian line fitting. From top to bottom, parameters for the light, normal and heavy Cs mass systems.}
 \begin{tabular}{|p{15mm}|p{15mm}|p{12mm}|p{1cm}|p{1cm}|p{12mm}|}\hline
 $\mathbf{q}=[x 0 0]$&$\kappa_{\rm P}^{\alpha{}=1}(\mathbf{q})$&$\omega$ (THz)&$\tau$ (ps)&$n$&$\nabla_{\mathbf{q}}\omega(\mathbf{q})$\\ \hline
 1&0.15702&0.191&78.1&1.000&0.166\\
2&0.23864&0.357&36.8&1.000&0.160\\
3&0.01539&0.510&23.4&0.129&0.132\\
4&0.01160&0.621&32.4&0.129&0.080\\
5&0.00235&0.670&15.0&0.129&0.049\\
\hline

1&0.19552&0.177&125&1.000&0.158\\
2&0.33325&0.335&61.8&1.000&0.155\\
3&0.00631&0.487&10.9&0.107&0.144\\
4&0.00214&0.622&8.4&0.050&0.112\\
5&0.00140&0.711&5.8&0.057&0.089\\
\hline
1&0.00025&0.102&89.4&0.048&0.073\\
2&0.00030&0.175&61.4&0.048&0.057\\
3&0.01473&0.215&78.9&0.048&0.285\\
4&0.02169&0.744&10.5&0.048&0.274\\
5&0.00008&0.762&11.1&0.038&0.018\\\hline
 \end{tabular}

 \label{table:param}
 \end{table}

\newpage 
\providecommand{\latin}[1]{#1}
\makeatletter
\providecommand{\doi}
  {\begingroup\let\do\@makeother\dospecials
  \catcode`\{=1 \catcode`\}=2 \doi@aux}
\providecommand{\doi@aux}[1]{\endgroup\texttt{#1}}
\makeatother
\providecommand*\mcitethebibliography{\thebibliography}
\csname @ifundefined\endcsname{endmcitethebibliography}
  {\let\endmcitethebibliography\endthebibliography}{}
